\documentclass[aoas,MSNbibl,nameyear,seceqn,dvips]{arximspdf}
\usepackage{subeqn}
\usepackage{graphicx}

\doi{10.1214/11-AOAS466}
\volume{5}
\issue{3}
\pubyear{2011}
\firstpage{1752}
\lastpage{1779}

\makeatletter
\renewcommand{\epsilon}{\varepsilon}

\def\bptnote#1{}

\def\@bmisc[#1]{%
  \get@battribute{unstr}%
  \common@pub@types%
  \let\bauthor\bbl@bauthor%
  \let\bhowpublished\@firstofone%
  \def\borganization##1{{\bauthor@style ##1}}%
}

\makeatother

\begin{document}
\begin{frontmatter}

\title{Measuring reproducibility of high-throughput experiments\thanksref{TT1}}
\runtitle{Reproducibility of high-throughput experiments}

\begin{aug}
\author[A]{\fnms{Qunhua}~\snm{Li}\corref{}\ead[label=e1]{qli@stat.berkeley.edu}},
\author[A]{\fnms{James~B.}~\snm{Brown}\ead[label=e2]{ben@newton.berkeley.edu}},
\author[A]{\fnms{Haiyan}~\snm{Huang}\ead[label=e3]{hhuang@stat.berkeley.edu}}
\and
\author[A]{\fnms{Peter~J.}~\snm{Bickel}\ead[label=e4]{bickel@stat.berkeley.edu}}
\runauthor{Li, Brown, Huang and Bickel}
\affiliation{University of California at Berkeley}
\address[A]{Department of Statistics\\
University of California\\
367 Evans Hall, Mail Stop 3860\\
Berkeley, California 94720\\
USA\\
\printead{e1}\\
\hphantom{\textsc{E-mail}: }\printead*{e2}\\
\hphantom{\textsc{E-mail}: }\printead*{e3}\\
\hphantom{\textsc{E-mail}: }\printead*{e4}}
\end{aug}
\thankstext{TT1}{Supported in part by NIH 1U01HG004695-01,
 1-RC2-HG005639-01 and  R21EY019094.}

% HISTORY:
\received{\smonth{5} \syear{2010}}
\revised{\smonth{1} \syear{2011}}

% ABSTRACT
%
\begin{abstract}
Reproducibility is essential to reliable scientific discovery in
high-throughput experiments. In this work we propose a unified approach
to measure the reproducibility of findings identified from replicate
experiments and identify putative discoveries using reproducibility.
Unlike the usual scalar measures of reproducibility, our approach
creates a curve, which quantitatively assesses when the findings are no
longer consistent across replicates. Our curve is fitted by a copula
mixture model, from which we derive a quantitative reproducibility
score, which we call the ``irreproducible discovery rate'' (IDR)
analogous to the FDR. This score can be computed at each set of paired
replicate ranks and permits the principled setting of thresholds both
for assessing reproducibility and combining replicates.

Since our approach permits an arbitrary scale for each replicate, it
provides useful descriptive measures in a wide variety of situations to
be explored. We study the performance of the algorithm using
simulations and give a heuristic analysis of its theoretical
properties. We demonstrate the effectiveness of our method in a
ChIP-seq experiment.\vspace*{-3pt}
\end{abstract}

% KEYWORDS
%
\begin{keyword}
\kwd{Reproducibility}
\kwd{association}
\kwd{mixture model}
\kwd{copula}
\kwd{iterative algorithm}
\kwd{irreproducible discovery rate}
\kwd{high-throughput experiment}
\kwd{genomics}.
\end{keyword}

\end{frontmatter}

%s1 ###
%se1 #&#
\section{Introduction}\label{sec1}

High-throughput profiling technologies play an indispensable role in
modern biology. By studying a large number of candidates in a single
experiment and assessing their significance using data analytical
tools, high-throughput technologies allow researchers to effectively
select potential targets for further studies. Despite their ubiquitous
presence in biological research, it is known that any single
experimental output from a high-throughput assay is often subject to
substantial variability. Reproducibility of high-throughput assays,
such as the level of agreement between results from replicate
experiments across (biological or technical)\vadjust{\eject} replicate samples, test
sites or experimental or data analytical platforms, is a constant
concern in their scientific applications [e.g., \citet{maqc2006} in
microarray experiments, \citet{park2009} in ChIP-seq technology].
Metrics that objectively assess the reproducibility of high-thoughput
assays are important for producing reliable scientific discoveries and
monitoring the performances of data generating procedures.

An important criterion for assessing reproducibility of results from
high-throughput experiments is how reproducibly top ranked signals are
reported in replicate experiments. These signals and their significance
scores, often presented as the primary results to be accessed by
downstream steps, are critical for prioritizing follow-up studies.
A common approach to assess this reproducibility is to compute the
Spearman's pairwise rank correlation coefficient between the
significance scores for signals that pass a prespecified significance
threshold on each replicate [see \citet{maqc2006} and \citet{kuo2006}
for examples in microarray studies].
However, the Spearman's correlation coefficient actually is not
entirely suitable for measuring the reproducibility between two rankings in
this type of application. First, this summary depends on the choice of
significance thresholds and may render false assessments that reflect
the effect of thresholds rather than the data generating procedure to
be evaluated. For instance, with everything else being equal, stringent
thresholds generally produce higher rank correlations than relaxed
thresholds when applied to the same data.
Although standardizing thresholds in principle can remove this
confounding effect, calibration of scoring systems across replicate
samples or different methods is challenging in practice, especially
when the scores or their scales are incomparable on replicate outputs.
Though this difficulty seemingly is associated only with
heuristics-based scores, indeed, it is also present for
probabilistic-based scores, such as $p$-values, if the probabilistic
model is ill-defined. For example, it has been reported in large-scale
systematic analyses that strict reliance on $p$-values in reporting
differentially expressed genes causes an apparent lack of
inter-platform reproducibility in microarray experiments [\citet{maqc2006}].
Second, the rank correlation treats all ranks equally, though the
differences in the top ranks seem to be more critical for judging the
reproducibility of findings from high-throughput experiments.
Alternative measures of correlation that give more importance to higher
ranks than lower ones, for instance, by weighing the difference of
ranks differently, have been developed in more general settings [e.g.,
\citet{blest2000}; \citet{genest2003b}; \citet{dacosta2005}] and applied to this
application [see \citet{boulesteix2009} for a review]. However, all
these measures are also subject to prespecified thresholds and raise
the question of how to select the optimal weighing scheme.

In this work we take an alternative approach to measure the
reproducibility of results in replicate experiments.
Instead of depending on a prespecified threshold, we describe
reproducibility as the extent to which the ranks\vadjust{\eject} of the signals are no
longer consistent across replicates in decreasing significance.
We propose a copula-based graphical tool to visualize the loss of
consistency and localize the possible breakdown of consistency
empirically. We further quantify reproducibility by classifying signals
into a reproducible and an irreproducible group, using a copula mixture
model. By jointly modeling the significance of scores on individual
replicates and their consistency between replicates, our model assigns
each signal a reproducibility index,
which estimates its probability to be reproducible.
Based on this index, we then define the irreproducible discovery rate
(IDR) and a selection procedure, in a fashion analogous to their
counterparts in multiple testing,\vadjust{\goodbreak} to rank and select signals.
As we will illustrate, the selection by this reproducibility criterion
provides the potential for more accurate classification. The overall
reproducibility of the replicates is described using IDR as the average
amount of irreproducibility in the signals selected.

The proposed approach, indeed, is a general method that can be applied
to any ranking systems that produce scores without ties, though we
discuss it in the context of high-throughput experiments. Because our
copula-based approach does not make any parametric assumptions on the
marginal distributions of scores, it is applicable to
both probabilistic- and heuristic-based scores.
When a threshold is difficult to determine in a scoring system, for
example, heuristic-based scores, it provides a reproducibility-based
criterion for setting selection thresholds.

In the next section we present the proposed graphical tool (Section \ref{SSprofile}),
the copula mixture model and its estimation procedure
(Section \ref{SSinference}), and the reproducibility criterion
(Section \ref{SSidr}). In Section \ref{Ssimulation} we use
simulations to evaluate the performance of our model, and compare with
some existing methods. In Section \ref{Sencode} we apply our method to
a data set that motivated this work. The data set was generated by the
ENCODE consortium [{\citet{encode2004}] from a ChIP-seq assay, a
high-throughput technology for studying protein-binding regions on the
genome. The primary interest is to assess the reproducibility of
several commonly used and publicly available algorithms for identifying
the protein-binding regions in ChIP-seq data. Using this data, we
compare the reproducibility of these algorithms in replicate
experiments, infer the reliability of signals identified by each
algorithm, and demonstrate how to use our method to identify suboptimal
results. Section \ref{Sdiscussion} is a general discussion. Finally,
we present a heuristic justification of our algorithm on optimality
grounds in the supplementary materials [\citet{li2011}].

%s2 ###
%se2 #&#
\section{Statistical methods}\label{sec2}

The data that we consider consist of a large number of putative signals
measured on very few replicates of the same underlying stochastic
process, for example, protein binding sites identified on the genomes
of biological replicates in ChIP-seq experiments. We assume that each
putative signal has been assigned a score that relates to the strength
of the evidence for the signal to be real on the corresponding
replicate by some data analysis method. The score can be either
heuristic based (e.g., fold enrichment) or probabilistic based (e.g.,
$p$-value). We further assume that all the signals are assigned distinct
significance scores and that the significance scores reasonably
represent the relative ranking of signals. However, the distribution
and the scale of the scores are unknown and can vary on different
replicates. We assume without loss of generality that high scores
represent strong evidence of being genuine signals and are ranked high.
By convention, we take the ``highest'' rank to be 1 and so on. We shall
use the scores as our data.

We assume $n$ putative signals are measured and reported on each replicate.
Then the data consist of $n$ signals on each of the $m$ replicates,
with the corresponding vector of scores for signal $i$ being $(x_{i,1},
\ldots, x_{i,m})$. Here $x_{i,j}$ is a scalar score for the signal on
replicate $j$.
Our goal is to measure the reproducibility of scores across replicates and
select reliable signals by considering information on the replicates
jointly. In what follows, we focus on the case of two replicates (i.e.,
$m=2$), although the methods in this paper can be extended to the case
with more replicates (see supplementary materials [\citet{li2011}]).

If replicates measure the same underlying stochastic process, then for
a reasonable scoring system, the significance scores of genuine signals
are expected to be ranked not only higher but also more consistently on
the replicates than those of spurious signals. When ranking signals by
their significance scores, a (high) positive association is expected
between the ranks of scores for genuine signals. A~degradation or a
breakdown of consistency between ranks may be observed when getting
into the noise level. This change of association provides an internal
indicator of the transition from real signal to noise. We will use this
information in measuring the reproducibility of signals.

In this section we first present a graphical tool (Section \ref{SSprofile}) for visualizing the change of association and localizing
the possible breakdown of association, empirically without model
assumptions. We then present a model-based approach (Section \ref{SSinference}), which quantifies the heterogeneity of association and
leads to a reproducibility criterion for threshold selection.

%s2.1 ###
%su2.1 #&#
\subsection{Displaying the change of association}\label{sec2.1}\label{SSprofile}

As we mentioned, the bivariate association between the significance
scores is expected to be positive for significant signals, then
transits to zero when getting into noise level. By visualizing how
association changes in the decreasing order of significance, one may
localize the transition of association and describe reproducibility in
terms of how soon consistency breaks down and how much empirical
consistency departs from perfect association before the breakdown.

Rank-based graphs are useful tools for displaying bivariate dependence
structure, because they are invariant with respect to monotone
transformations of the variables and are thus scale free. Earlier
papers have proposed rank-based graphical tools, such as the Chi-plot
[\citeauthor{fisher1985} (\citeyear{fisher1985,fisher2001})]
and the Kendall plot [\citet{genest2003}], for visualizing the presence of association in samples
from continuous bivariate distributions. Related to nonparametric tests
of independence, these graphs primarily are designed for detecting
bivariate dependence by representing the presence of association as
departures from the pattern under independence. The type and the level
of simple bivariate association may be inferred by comparing the
patterns of dependence observed in these plots with the prototypical
patterns in \citeauthor{fisher1985} (\citeyear{fisher1985,fisher2001}), \citet{genest2003}. However,
these graphs are less informative, when heterogeneity of association,
such as the one described here, is present. (See Figure \ref{Fsalmon}
in Section \ref{SSScurve-ex} for an illustration on a real data set
with mixed populations.)

We now present our rank-based graph, which we refer to as a
correspondence curve, intended to explicitly display the aforementioned
change of association.

%s2.1.1 ###
%su2.1.1 #&#
\subsubsection{Correspondence curves}\label{sec2.1.1}\label{SSScurve}

Let $(X_{1,1}, X_{1,2}), \ldots, (X_{n,1}, X_{n,2})$ be a sample of
scores of $n$ signals on a pair of replicates.
Define
%
%e2.1 ###
%e2.1 #&#
%e2.2 #&#
\begin{eqnarray}\label{EURI}
\Psi_n(t, v) = \frac{1}{n} \sum_{i=1}^{n} 1\bigl(X_{i,1} > x_{(\lceil
(1-t)n\rceil),1}, X_{i,2} > x_{(\lceil(1-v)n\rceil),2}\bigr),\\
  \eqntext{0 < t
\leq1, 0 < v \leq1,}
\end{eqnarray}
where $x_{(\lceil(1-t)n\rceil),1}$ and $x_{(\lceil(1-v)n\rceil),2}$
denote the order statistics of $X_1$ and $X_2$, respectively. $\Psi
_n(t, v)$ essentially describes the proportion of the pairs that are
ranked both on the upper $t\%$ of $X_1$ and on the upper $v\%$ of
$X_2$, that is, the intersection of upper ranked identifications. As
consistency usually is deemed a symmetric notion, we will just focus on
the special case of $t=v$ and use the shorthand notation $\Psi_n(t)$ in
what follows. In fact, $\Psi_n(t, v)$ is an empirical survival copula
[\citet{nelson1999}], and $\Psi_n(t)$ is the diagonal section of $\Psi
_n(t, v)$ [\citet{nelson1999}]. (See Section \ref{SSScopulas} for a
brief introduction of copulas.) Define the population version $\Psi(t)
\equiv\lim_n \Psi_n(t)$. Then $\Psi(t)$ and its derivative $\Psi'(t)$,
which represent the change of consistency, have the following
properties. (See supplementary materials [\citet{li2011}] for
derivation.)

Let $R(X_{i,1})$ and $R(X_{i,2})$ be the ranks of $X_{i,1}$ and
$X_{i,2}$, respectively.
\begin{longlist}[(3)]
\item[(1)] If $R(X_{i,1})=R(X_{i,2})$ for $X_{i,j} \in(F_j^{-1}(1-t),
F_j^{-1}(1-t_0)], j=1, 2$, with $0 \leq t_0 \leq t \leq1$, $\Psi(t) = \Psi
(t_0)+t - t_0$ and $\Psi'(t)=1$.
\item[(2)] If $R(X_{i,1}) \perp R(X_{i,2})$ for $X_{i,j} \in(F_j^{-1}(1-t),
F_j^{-1}(1)], j=1, 2$, with $0 \leq t \leq1$, $\Psi(t)=t^2$ and $\Psi'(t)=2t$.
\item[(3)] If $R(X_{i,1})=R(X_{i,2})$ for $X_{i,j} \in(F_j^{-1}(1-t_0),
F_j^{-1}(1)]$\vspace*{-1pt} and $R(X_1) \perp R(X_2)$ for $X_{i,j} \in(F_j^{-1}(0),
F_j^{-1}(1-t_0)], j=1, 2$, with\vspace*{-1pt} $0 \leq t_0 \leq1$, then for $t_0 \leq
t \leq1$, $\Psi(t) = \frac{t^2-2tt_0 + t_0}{1-t_0}$ and $\Psi'(t)
=\frac{2(t-t_0)}{1-t_0}$.
\end{longlist}

The last case describes an idealized situation in our applications,
where all the genuine signals are ranked higher than any spurious
signals, and the ranks on the replicates are perfectly correlated for
genuine signals but completely independent for spurious signals. The
same properties are approximately followed in the corresponding sample
version $\Psi_n$ and $\Psi_n'$ with finite differences replacing derivatives.

To visualize the change of consistency with the decrease of
significance, a curve can be constructed by plotting the pairs $(t, \Psi
_n(t))$ [or $(t, \Psi'_n(t))$] for $0 \leq t \leq1$. The resulting
graphs, which we will refer to as a correspondence curve (or a change
of correspondence curve, resp.), depend on $X_{i,1}$ and~$X_{i,2}$
only through their ranks, and are invariant to both location
and scale transformation on $X_{i,1}$ and $X_{i,2}$. Corresponding to
the three special cases described earlier, the curves have the
following patterns:
\begin{longlist}[(3)]
\item[(1)] When $R(X_{i,1})$ and $R(X_{i,2})$ are perfectly correlated for
$i=1, \ldots, n$, all points on the curve of $\Psi_n$ will fall on a
straight line of slope 1, and all points on the curve of $\Psi'_n$ will
fall on a straight line with slope 0.
\item[(2)] When $R(X_{i,1})$ and $R(X_{i,2})$ are independent for $i=1,
\ldots, n$, all points on the curve of $\Psi_n$ will fall on a parabola
$t^2$, and all points on the curve of~$\Psi_n'$ fall on a straight line
of slope of $2t$.
\item[(3)] When $R(X_{i,1})$ and $R(X_{i,2})$ are perfectly correlated for
the top $t_0n$ observations and independent for the remaining
$(1-t_0)n$, top $t_0n$ points fall into a straight line of slope 1 on
the curve of $\Psi_n$ and slope 0 on the curve of $\Psi_n'$,\vspace*{-1pt} and the
rest $(1-t_0)n$ points fall into a parabola $\Psi_n(t)=\frac{t^2-2tt_0
+ t_0}{1-t_0}$ ($t > t_0$) on the curve\vspace*{-1pt} of $\Psi_n$ and a straight line
of slope $\frac{2(t-t_0)}{1-t_0}$ on the curve of $\Psi_n'$.
\end{longlist}

These properties show that the level of positive association and the
possible change of association can be read off these types of curves.
For the curve of $\Psi_n$, strong association translates into a nearly
straight line of slope~1, and lack of association shows as departures
from the diagonal line, such as curvature bending toward the $x$-axis
[i.e., $\Psi_n(t) < t$]; if almost no association is present, the curve
shows a parabolic shape. Similarly, for the curve of~$\Psi_n'$, strong
association translates into a nearly straight line of slope 0, and lack
of association shows as a line with a positive slope. The transition of
the shape of the curves, if present, indicates the breakdown of
consistency, which provides guidance on when the signals become spurious.

%s2.1.2 ###
%su2.1.2 #&#
\subsubsection{Illustration of the correspondence curves}\label{sec2.1.2}\label{SSScurve-ex}

We first demonstrate the curves
using an idealized case (Figure \ref{Fideal}), where $R(X)$ and $R(Y)$
agree perfectly for the top $50\%$ of observations and are independent
for\vadjust{\eject} the remaining $50\%$ of observations. The curves display the
pattern described in case 3 above.
The transition of the shape of the curves occurs at $50\%$, which
corresponds to the occurence of the breakdown of consistency.
Transition can be seen more visibly on the curve of $\Psi_n'$ by the
gap between the disjoint lines with 0 and positive slopes, which makes\vadjust{\goodbreak} $\Psi_n'$ a better choice for inspecting and localizing
the transition than $\Psi_n$, especially when the transition is less sharp.
More simulated examples are presented in Section \ref{Ssimulation} to
illustrate the curves in the presence and absence of the transition of
association in more realistic settings.

%f1 ###
%fi1 #&#
\begin{figure}[t]

\includegraphics{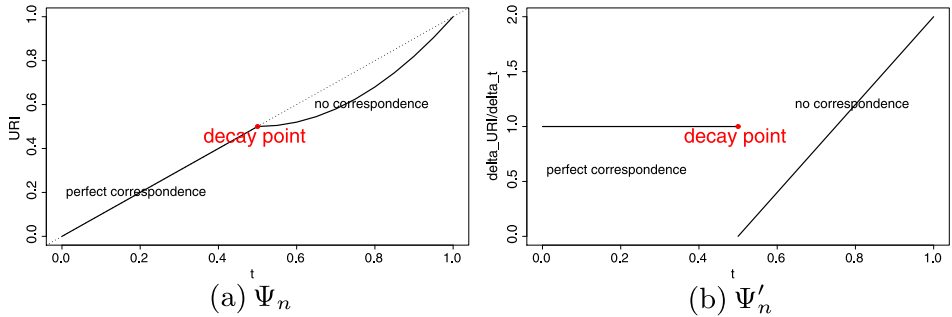}

\caption{An illustration of the correspondence profile in an idealized
case, where top 50\% are genuine signals and bottom 50\% are noise. In
this case, all signals are ranked higher than noise; two rank lists
have perfect correspondence for signals and no correspondence for
noise. \textup{(a)}  Correspondence curve. \textup{(b)}  Change of correspondence curve.}\label{Fideal}
\end{figure}

%f2 ###
%fi2 #&#
\begin{figure}[t]

\includegraphics{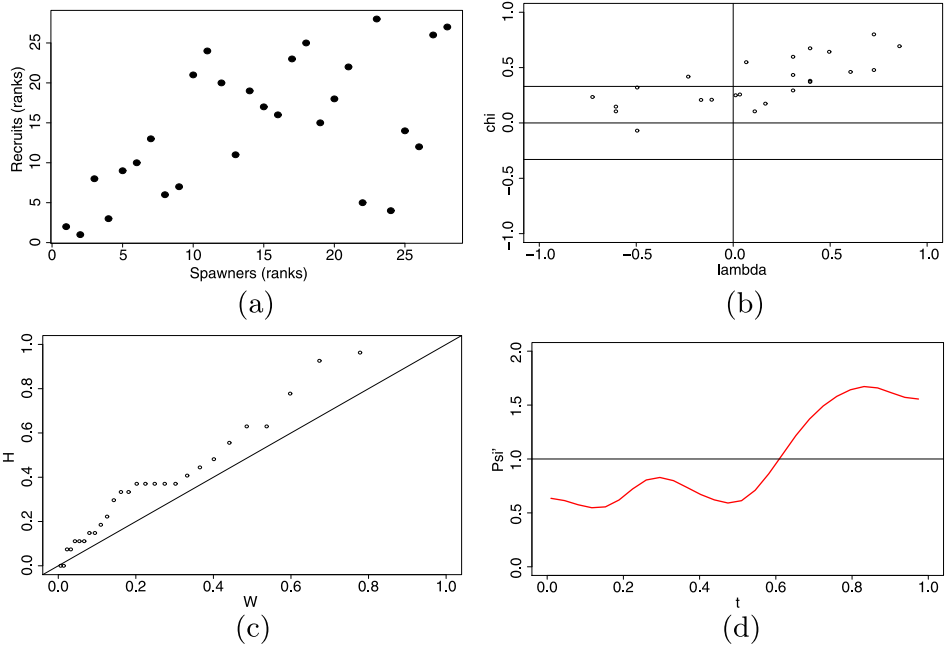}

\caption{Rank scatterplot \textup{(a)}, Chi-plot \textup{(b)}, $K$-plot \textup{(c)} and the change
of correspondence curve \textup{(d)} for salmon data, which consists of 28 measurements of size of the annual
spawning stock of salmon and corresponding production of new
catchable-sized fish in the Skeena River. The curve of $\Psi_n'$ is
produced by taking derivative on the spline that fits $\Psi_n$ with
df${}={}$6.4.}\label{Fsalmon}
\end{figure}

We now compare the $\Psi_n'$ plot with the Chi-plot and the $K$-plot
using a real example considered in \citet{kallenberg1999}, \citet{fisher2001},
\citet{genest2003}.
This data set consists of 28 measurements of size of the annual
spawning stock of salmon and corresponding production of new
catchable-sized fish in the Skeena River. It was speculated by \citet{fisher2001} to contain a mixed populations with heterogeneous
association. Though the dissimilarity of Chi-plot or $K$-plot to their
prototypical plots [cf. \citet{fisher2001}; \citet{genest2003}] suggests the
data may involve more than simple monotone association [\citet{fisher2001}; \citet{genest2003}],
neither of these plots manifest
heterogeneity of association.
In the $\Psi'_n$ curve [Figure~\ref{Fsalmon}(d)], the characteristic
pattern of transition is observed at about $t=0.5$, which indicates
that the data is likely to consist of two groups, with roughly the top
50\% from a strongly associated group and the bottom 50\% from a weakly
associated group. It agrees with the speculation in \citet{fisher2001}.

%s2.2 ###
%su2.2 #&#
\subsection{Inferring the reproducibility of signals}\label{sec2.2}\label{SSinference}

In this section we present a statistical model that quantifies the
dependence structure and infers the reliability of signals.
Throughout this section, we will suppose, for simplicity, that we are
dealing with a sample of i.i.d. observations from a population. Though
this is in fact unrealistic in many applications, in particular, for
the signals from genome-wide profiling (e.g., ChIP-seq experiments),
where observations are often dependent, the descriptive and graphical
value of our method remains, as we are concerned with first order effects.

In general, genuine signals tend to be more reproducible and score
higher than spurious ones. The scores on replicates may be viewed as a
mixture of two groups, which differ in both the strength of association
and the level of significance. Recall that in these applications, the
distributions and the scales of scores are usually unknown and may vary
across data sets.
To model such data, a semiparametric copula model is appropriate, in
which the associations among the variables are represented by a simple
parametric model but the marginal distributions are estimated
nonparametrically using their ranks to permit arbitrary scales. Though
using ranks, instead of the raw values of scores, generally causes some
loss of information, this loss is known to be asymptotically negligible
[\citet{lehmann2006}].
In view of the heterogeneous association in the genuine and spurious
signals, we further model the heterogeneity of the dependence structure
in the copula model using a mixture model framework.

Before proceeding to our model, we first provide a brief review of
copula models, and refer to \citet{joe1997} and \citet{nelson1999}
for a modern treatment of copula theory.

%s2.2.1 ###
%su2.2.1 #&#
\subsubsection{Copulas}\label{sec2.2.1}\label{SSScopulas}

The multivariate function $C=C(u_1, \ldots, u_p)$ is called a~copula
if it is a continuous distribution function and each marginal is a
uniform distribution function on $[0, 1]$. That is, $C\dvtx [0, 1]^p
\rightarrow[0, 1]$, with
$C(u)=P(U_1 \leq u_1, \ldots , U_p \leq u_p)$,
in which each $U_j \sim \operatorname{Unif}[0, 1]$ and $u=(u_1, \ldots , u_p)$.
By Sklar's theorem [\citet{sklar1959}], every continuous multivariate
probability distribution can be represented by its univariate marginal
distributions and a copula, described using a bivariate case as follows.

Let $X_1$ and $X_2$ be two random variables with continuous CDFs $F_1$
and~$F_2$. The copula $C$ of $X_1$ and $X_2$ can be found by making the
marginal probability integral transforms on $X_1$ and $X_2$ so that
%
%e2.2 ###
%e2.3 #&#
\begin{equation}\label{Ecopula}
C(u_1, u_2) =F(F_1^{-1}(u_1), F_2^{-1}(u_2)), \qquad u_1, u_2 \in[0, 1],
\end{equation}
where $F$ is the joint distribution function of $(X_1, X_2)$, $F_1$ and
$F_2$ are the marginal distribution functions of $X_1$ and $X_2$,
respectively, and $F_1^{-1}$ and~$F_2^{-1}$ are the right-continuous
inverses of $F_1$ and $F_2$, defined as $F_j^{-1}(u)=\operatorname{inf}\{
z\dvtx\break F_j(z) \geq u\}$. That is, the copula is the joint distribution of
$F_1(X_1)$, $F_2(X_2)$. These variables are unobservable but estimable
by the normalized ranks $F_{n1}(X_1)$, $F_{n2}(X_2)$ where $F_{n1}$,
$F_{n2}$ are the empirical distribution functions of the sample. The
function $\delta_C(t, t)= C(t, t)$ is usually referred to as the
diagonal section of a copula $C$. We will use the survival function of
the copula~$C$, $\bar{C}(u_1, u_2)=P(U_1>1-u_1, U_2>1-u_2)$, which
describes the relationship between the joint survival function [$\bar
{F}(x_1, x_2)=P(X_1>x_1, X_2>x_2)$] and its univariate margins ($\bar
{F}_j=1-F_j$) in a manner completely analogous to the relationship
between univariate and joint functions, as $\bar{C}(u_1, u_2)=\bar
{F}(\bar{F}_1^{-1}(u_1), \bar{F}_2^{-1}(u_2))$.
The sample version of (\ref{Ecopula}) is called an \textit{empirical
copula} [\citet{deheuvels1979}; \citet{nelson1999}], defined as
%
%e2.3 ###
%e2.4 #&#
\begin{equation}
C_n\biggl(\frac{i}{n}, \frac{j}{n}\biggr)= \frac{1}{n}\sum_{k=1}^n 1\bigl(x_{k, 1} \leq
x_{(i), 1}, x_{k, 2} \leq x_{(j), 2}\bigr), \qquad1 \leq i, j \leq n,
\end{equation}
for a sample of size $n$, where $x_{(i), 1}$ and $x_{(j), 2}$ denote
order statistics on each coordinate from the sample. The sample version
of survival copulas follows similarly.

This representation provides a way to parametrize the dependence
structure between random variables separately from the marginal
distributions, for example, a~parametric model for the joint
distribution of $u_1$ and $u_2$ and a nonparametric model for marginals.
Copula-based\vadjust{\eject} models are natural in situations where learning about the
association between the variables is important, but the marginal
distributions are assumably unknown.
For example, the 2-dimensional Gaussian copula $C$ is defined as
%
%e2.4 ###
%e2.5 #&#
\begin{equation}\label{Egaussian}
C(u_1, u_2 |\rho)=\Phi_2(\Phi^{-1}(u_1), \Phi^{-1}(u_2)|\rho),
\end{equation}
where $\Phi$ is the standard normal cumulative distribution function,
$\Phi_2(\cdot, \cdot|\rho)$ is the cumulative distribution function
for a bivariate normal vector
$(z_1, z_2) \sim N\left(\left(0 \atop 0\right),  \left(
{ 1 \atop \rho}\enskip{ \rho\atop 1
}
 \right)\right)$, and $\rho$ is the correlation coefficient. Modeling
dependence with arbitrary marginals $F_1$ and $F_2$ using the Gaussian
copula (\ref{Egaussian}) amounts to assuming data is generated from
latent variables $(z_1, z_2)$ by setting $x_1= F_1^{-1}(\Phi(z_1))$ and
$x_2= F_2^{-1}(\Phi(z_2))$.
Note that if $F_1$ and $F_2$ are not continuous, $u_1$ and $u_2$ are
not uniform. For convenience, we assume that $F_1$ and $F_2$ are
continuous throughout the text.

%s2.2.2 ###
%su2.2.2 #&#
\subsubsection{A copula mixture model}\label{sec2.2.2}\label{SSSmixture}

We now present our model for quantifying the dependence structure and
inferring the reproducibility of signals. We assume throughout this
part that our data is a sample of i.i.d. bivariate vectors $(x_{i, 1},
x_{i, 2})$.

We assume the data consists of genuine signals and spurious signals,
which in general correspond to a more reproducible group and a less
reproducible group.
We use the indicator $K_i$ to represent whether a signal $i$ is genuine
($K_i=1$) or spurious ($K_i=0$). Let $\pi_1$ and $\pi_0=1-\pi_1$ denote
the proportion of genuine and spurious signals, respectively. Given
$K_i=1$, we assume the pairs of scores for genuine (resp.,
spurious) signals are independent draws from a continuous bivariate
distribution with density $f_1(\cdot, \cdot)$ [resp., $f_0(\cdot
, \cdot)$, given $K_i=0$] with joint distribution $F_1(\cdot,\cdot)$
[resp., $F_0(\cdot, \cdot)$]. Note, however, that even if the
marginal scales are known, $K_i$ would be unobservable so that the
copula is generated by the marginal mixture (with respect to $K_i$),
$F_j =\pi_0F_j^0 + \pi_1F_j^1$, where $F_j$ is the\vspace*{-1pt} marginal
distribution of the $j${{th}} coordinate and $F_j^k$ is the
marginal distribution of the corresponding $k${{th}} component.

Because genuine signals generally are more significant and more
reproducible than spurious signals, we expect the two groups to have
both different means and different dependence structures between replicates.
We assume that, given the indicator $K_i$, the dependence between
replicates for genuine (resp., spurious) signals is induced by a
bivariate Gaussian distribution $\mathbf{z}_{1}=(z_{1, 1}, z_{1, 2})$
[or resp., $\mathbf{z}_{0}=(z_{0, 1}, z_{0, 2})$]. The choice of
Gaussian distribution for inducing the dependence structure in each
component is made based on the observation that the dependence within a
component in the data we consider generally is symmetric and that an
association parameter with a simple interpretation, such as the
correlation coefficient for a~Gaussian distribution, is natural.

As the scores from $F_1(\cdot, \cdot)$ are expected to be higher than
the scores from~$F_0(\cdot, \cdot)$,
we assume $\mathbf{z}_1$ has a higher mean than $\mathbf{z}_0$.
Since spurious signals are presumably less reproducible, we assume
corresponding signals on\vadjust{\eject} the replicates to be independent, that is,
$\rho_0=0$; whereas, since genuine signals usually are positively
associated between replicates, we expect $\rho_1 > 0$, though $\rho_1$
is not required to be positive in our model.
It also seems natural to assume that the underlying latent variables,
reflecting replicates, have the same marginal distributions. Finally,
we note that if the marginal scales are unknown, we can only identify
the \textit{difference} in means of the two latent variables and the \textit{ratio}
of their variances, but not the means and variances of the
latent variables. Thus, the parametric model generating our copula can
be described as follows:

Let $K_i \sim \operatorname{Bernoulli}(\pi_1)$ and $(z_{i,1}, z_{i,2})$ be distributed as
\begin{subequations}\label{Emixture}
%
%e2.5 ###
%e2.6 #&#
\begin{equation}\label{Estage1}
\pmatrix{z_{i,1} \cr z_{i,2}} \Bigm| K_i=k \sim N \left(\pmatrix{\mu_k \cr\mu_k},
 \pmatrix{\displaystyle
\sigma_k^2 &\rho_k\sigma_k^2 \vspace*{1pt}\cr\displaystyle
\rho_k\sigma_k^2 & \sigma_k^2
}
\right ),
\qquad k=0, 1,
\end{equation}
 where $\mu_0=0$, $\mu_1>0$, $\sigma_0^2=1$, $\rho_0=0$, $0< \rho
_1 \leq1$.

 Let
%
%e2.6 ###
%e2.7 #&#
\begin{eqnarray}\label{Estage2}
u_{i, 1} &\equiv& G(z_{i,1})= \frac{\pi_1}{\sigma_1}\Phi\biggl(\frac{z_{i,1} -
\mu_1}{\sigma_1}\biggr)+\pi_0\Phi(z_{i,1}), \nonumber
\\[-8pt]
\\[-8pt]
u_{i, 2} &\equiv& G(z_{i,2})= \frac{\pi_1}{\sigma_1}\Phi\biggl(\frac{z_{i,2} -
\mu_1}{\sigma_1}\biggr)+\pi_0\Phi(z_{i,2}).
\nonumber
\end{eqnarray}
 Our actual observations are
%
%e2.7 ###
%e2.8 #&#
\begin{eqnarray}\label{Estage3}
x_{i,1} &=& F_1^{-1}(u_{i,1}), \nonumber
\\[-8pt]
\\[-8pt]
x_{i,2} &=& F_2^{-1}(u_{i,2}),
\nonumber
\end{eqnarray}
\end{subequations}
\noindent
where $F_1$ and $F_2$ are the marginal distributions of the two
coordinates, which are assumed continuous but otherwise unknown.

Thus, our model, which we shall call a copula mixture model, is a
semiparametric model parametrized by $\theta=(\pi_1, \mu_1, \sigma_1^2,
\rho_1)$ and $(F_1, F_2)$.
The corresponding mixture likelihood for the data is
\begin{subequations}\label{ELmixture}
%
%e2.8 ###
%e2.9 #&#
%e2.10 #&#
\begin{eqnarray}\label{ELmixture1}
L(\theta)
&=&\prod_{i=1}^{n}[\pi_0 h_0(G^{-1}(F_1(x_{i,1})),
G^{-1}(F_2(x_{i,2})))\nonumber\\
&&\hphantom{\prod_{i=1}^{n}[}{}
+\pi_1 h_1(G^{-1}(F_1(x_{i,1})), G^{-1}(F_2(x_{i,2})))]
\\\label{ELmixture2}
&=& \prod_{i=1}^{n}[ c(F_1(x_{i,1}),
F_2(x_{i,2}))g(G^{-1}(F_1(x_{i,1})))g(G^{-1}(F_2(x_{i,2})))],
\end{eqnarray}
\end{subequations}
\setcounter{equation}{6}
\noindent
where
%
%e2.9 ###
%e2.7 #&#
\begin{equation}
c(u_1, u_2) = \frac{\pi_0 h_0(G^{-1}(u_1), G^{-1}(u_2))+\pi_1
h_1(G^{-1}(u_1), G^{-1}(u_2))}{g(G^{-1}(u_1))g(G^{-1}(u_2))}
\end{equation}
is a copula density function with
\[
h_0 \sim N \left( \left(\matrix{0 \cr 0}\right),   \left(\matrix{
  1& 0  \cr 0 & 1
}
\right )  \right )   \quad \mbox{and} \quad
  h_1 \sim  N  \left(\left(\matrix{ \mu_1 \cr\mu_1}\right), \left(\matrix{
  \sigma_1^2 & \rho_1\sigma_1^2 \vspace*{1pt}\cr\rho_1\sigma_1^2 &
\sigma_1^2}
 \right) \right ),
 \]
$G$ is defined in (\ref{Estage2}) and $g$ is the density function of
$G$. Note that $G$ depends on~$\theta$.

Given the parameters $\theta$, the posterior probability that a signal
$i$ is in the irreproducible group can be computed as
%
%e2.10 ###
%e2.8 #&#
\begin{equation}\label{Eposterior}
 \qquad
\operatorname{Pr}\bigl(K_i=0 \mid(x_{i1}, x_{i2}); \theta\bigr) =
\frac{\pi_0
h_0(G^{-1}(F_1(x_{i,1}), G^{-1}(F_2(x_{i,2}))))}{\sum_{k=0,1}\pi_k
h_k(G^{-1}(F_1(x_{i,1}), G^{-1}(F_2(x_{i,2}))))}.\hspace*{-5pt}
\end{equation}
We estimate values for these classification probabilities by estimating
the parameters $\theta$ using an estimation procedure described in
Section \ref{SSSestimation}, and substituting these estimates into the
above formulas.

The idea of using a mixture of copulas to describe complex dependence
structures is not entirely new. For example, the mixed copula model
[\citet{hu2006}] in economics uses
a mixture of copulas [$C_{\mathit{mix}}(u_1, u_2 \mid(\theta_1, \ldots, \theta
_k))= \sum_{i=1}^k C(u_1, u_2 \mid\theta_i)$]
to generate flexible fits to the dependence structures that do not
follow any standard copula families.
In this model, all the copulas in $C_{\mathit{mix}}$ are assumed to have
identical marginal distributions.
In contrast, the copula in our model not only has mixed associations,
but also allows different associations to occur with different marginal
distributions ($F_j^0$ and $F_j^1$), thus can be viewed as a
generalization of the case with the same marginal distribution. In
addition, our modeling goal is to cluster the observations into groups
with homogeneous associations, instead of data fitting. This difference
in marginal distributions calls for nonstandard estimation, which we
expect to be efficient,
as we shall see in Section \ref{SSSestimation}.

%s2.2.3 ###
%su2.2.3 #&#
\subsubsection{Estimation of the copula mixture model}\label{sec2.2.3}\label{SSSestimation}

In this section we describe an estimation procedure that estimates the
parameters $\theta$ in (\ref{ELmixture}) and the membership $K_i$ of
each observation.

A common strategy to estimate the association parameters in
semiparametric copula models is a ``pseudo-likelihood'' approach, which
is described in broad, nontechnical terms by \citet{oakes1994}.
In this approach, the empirical marginal distribution functions $\hat
{F}_{j}$, after rescaling by multiplying by ($\frac{n}{n+1}$) to avoid
infinities, are plugged into the copula density in (\ref{ELmixture2}),
ignoring the terms involving $g$. The association parameters are then
estimated by maximizing the pseudo-copula likelihood.  \citet{genest1995}  showed, without specifying the
algorithms to compute them, that under certain technical conditions,
the estimators obtained from this approach are consistent,
asymptotically normal, and fully efficient only if the coordinates of
the copula are independent.

We adopt a different approach which, in principle, leads to efficient
estimators under any choice of parameters and $F_1$, $F_2$. Note that
the estimation\vadjust{\eject} of the association parameter $\rho_1$ depends on the
estimation of $\mu_1, \sigma_1^2$ and $\pi_1$ due to the presence of
the mixture structure on marginal distributions, which makes the
log-likelihood (\ref{ELmixture}) difficult to maximize directly. Our
approach is to estimate\vspace*{1pt} the parameters $\hat{\theta}$ by maximizing the
log-likelihood (\ref{ELmixture}) of pseudo-data $G^{-1}(\frac
{n}{n+1}\hat{F}_{i,j}; \theta)$,\vspace*{-2pt} where $\hat{F}_{i,j} \equiv\hat
{F}_j(x_{i,j})$.

As the latent variables $z_{0, j}$ and $z_{1, j}$ in our model form a
mixture distribution, it is natural to use an expectation--maximization
(EM) algorithm [\citet{dempster1977}] to estimate the parameters $\hat
{\theta}$ and infer the status of each putative signal for pseudo-data.
In our approach, we first compute the pseudo-data $G^{-1}(\frac
{n}{n+1}\hat{F}_{i,j}; \theta_0)$ from some initialization parameters
$\theta^{(0)}$, then iterate between two stages: (1) maximizing $\theta
$ based on the pseudo-data using EM and (2) updating the pseudo-data.
The detailed procedure is given in the supplementary materials [\citet{li2011}].
The EM stage may be trapped in local maxima, and the stage
of updating pseudo-data may not converge from all starting points.
However, in the simulations we performed (Section \ref{Ssimulation}),
it behaves well and finds the global maxima, when started from a number
of initial points.

We sketch in the supplementary materials [\citet{li2011}, Section 2]
a~heuristic argument that a limit point of our algorithm close to the
true value satisfies an equation whose solution is asymptotically
efficient. Although our algorithm converges in practice, we have yet to
show its convergence in theory. However, a modification which we are
investigating does converge to the fixed point mentioned above. This
work will appear elsewhere.\looseness=-1

%s2.3 ###
%su2.3 #&#
\subsection{Irreproducible identification rate}\label{sec2.3}\label{SSidr}

In this section we derive a reproducibility criterion from the copula
mixture model in Section \ref{SSSmixture} based on an analogy between
our method and the multiple hypothesis testing problem.\ This criterion
can be used to assess the reproducibility of both individual signals
and the overall replicate outputs.

In the multiple hypothesis testing literature, the false discovery rate
(FDR) and its variants, including positive false discovery rate (pFDR)
and marginal false discovery rate (mFDR), are introduced to control the
number of false positives in the rejected hypotheses [\citet{benjamini1995};
\citet{storey2002}; \citet{genovese2002}].
In the FDR context, when hypotheses are independent and identical, the
test statistics can be viewed as following a mixture distribution of
two classes, corresponding to whether or not the statistic is generated
according to the null hypothesis [e.g., \citet{efron2004a}; \citet{storey2002}].
Based on this mixture model, the local false discovery
rate, which is the posterior probability of being in the null component
$\operatorname{Lfdr}(\cdot)=(1-\pi)f_0(\cdot)/f(\cdot)$, was introduced to compute the
individual significance level [\citet{efron2004a}]. \citet{sun2007}
show, again for the i.i.d. case, that Lfdr is also an optimal statistic in
the sense that the thresholding rule based on Lfdr controls the
marginal false discovery rate with the minimum marginal false
nondiscovery rate.

As in multiple hypothesis testing, we also build our approach on a
mixture model and classify the observations into two classes. However,
the two classes have different interpretation and representation:
The two classes represent irreproducible measurements and reproducible
measurements in our model, in contrast to nulls and nonnulls in the
multiple testing context, respectively.

In analogy to the local false discovery rate, we define a quantity,
which we call the \textit{local irreproducible discovery rate}, to be
%
%e2.11 ###
%e2.9 #&#
\begin{equation}
 \operatorname{idr}(x_{i,1}, x_{i,2}) = \frac{\pi_0 h_0(G^{-1}(F_1(x_{i,1})),
G^{-1}(F_2(x_{i, 2})))}{\sum_{k=0,1} \pi_k h_k(G^{-1}(F_1(x_{i,1})),
G^{-1}(F_2(x_{i, 2})))}.
\end{equation}
This quantity can be thought of as the a posteriori probability
that a signal is not reproducible on a pair of replicates [i.e., (\ref{Eposterior})], and can be estimated from the copula mixture model.

Similarly, we define the \textit{irreproducible discovery rate} (IDR) in
analogy to the mFDR,
%
%e2.12 ###
%e2.10 #&#
\begin{eqnarray}\label{EIDR}
\operatorname{IDR}(\gamma) &=& P(\mathit{irreproducible} \mid i \in
I_{\gamma})\nonumber
\\[-8pt]
\\[-8pt]
&=& \frac{\pi_0 \int_{I_{\gamma}} dH_0(G^{-1}(F_1(x_{i,1})),
G^{-1}(F_2(x_{i,2})))}{\int_{I_{\gamma}}dH(G^{-1}(F_1(x_{i,1})),
G^{-1}(F_2(x_{i,2})))},
\nonumber
\end{eqnarray}
where $I_{\gamma} = \{(x_{i,1}, x_{i, 2})\dvtx  \operatorname{idr}(x_{i,1}, x_{i,2}) <
\gamma\}$, $H_0$ and $H$ are the CDF of density functions $h_0$ and
$h=\pi_0h_0 + \pi_1h_1$, respectively.
For a desired control level $\alpha$, if $(x_{(i), 1}, x_{(i),2})$ are the
pairs ranked by $\mathit{idr}$ values, define $l= \max\{i\dvtx  \frac{1}{i}\sum
_{j=1}^i \mathit{idr}_{j} \leq\alpha\}$. By selecting all $(x_{(i), 1},
x_{(i),2})$ ($i= 1, \ldots, l$), we can think of this procedure as
giving an expected rate of irreproducible discoveries no greater than
$\alpha$. It is analogous to the adaptive step-up procedure of \citet{sun2007}
for the multiple testing case.

This procedure essentially amounts to re-ranking the identifications
according to the likelihood ratio of the joint distribution of the two
replicates. The resulting rankings are generally different from the
ranking of the original significance scores on either replicate.

Unlike the multiple testing procedure, our procedure does not require
$x_{i,j}$ to be $p$-values; instead, $x_{i,j}$ can be any scores with
continuous marginal distributions. When $p$-values are used as scores,
our method can also be viewed as a method to combine $p$-values.
We compare our method and two commonly-used $p$-value combinations
through simulations in Sec\-tion~\ref{Ssimulation}.

%s3 ###
%se3 #&#
\section{Simulation studies}\label{sec3}\label{Ssimulation}

%s3.1 ###
%su3.1 #&#
\subsection{Illustration of correspondence curves}\label{sec3.1}

To show the prototypical plots of more realistic cases, we use
simulated data to compare and contrast the curves in presence and
absence of the transition of association described\vadjust{\eject} in Section \ref{SSprofile}.
(Figure \ref{Fcurve}). The case where no transition occurs is
illustrated using two single-component bivariate Gaussian distributions
with homogeneous association, $\rho=0$ [Figure \ref{Fcurve}(a)] and $\rho
=0.8$ [Figure \ref{Fcurve}(b)], respectively. The presence of the
transition is illustrated using two two-component bivariate Gaussian
mixtures, whose lower ranked component has independent coordinates
(i.e., $\rho_0=0$) and the higher ranked component has positively
correlated coordinates with $\rho_1=1$ [Figure \ref{Fcurve}(c)] and $\rho
_1=0.8$ [Figure \ref{Fcurve}(d)], respectively.

As in the idealized example (Figure \ref{Fideal}),
the characteristic transition of curves is observed when the
transition of association is present [Figure \ref{Fcurve}(c), (d)], but
not seen
when the data consists of only one component with homogeneous
association. This shows that the transition of the shape of the curve
may be used as an indicator for the presence of the transition of
association.

%f3 ###
%fi3 #&#
\begin{figure}

\includegraphics{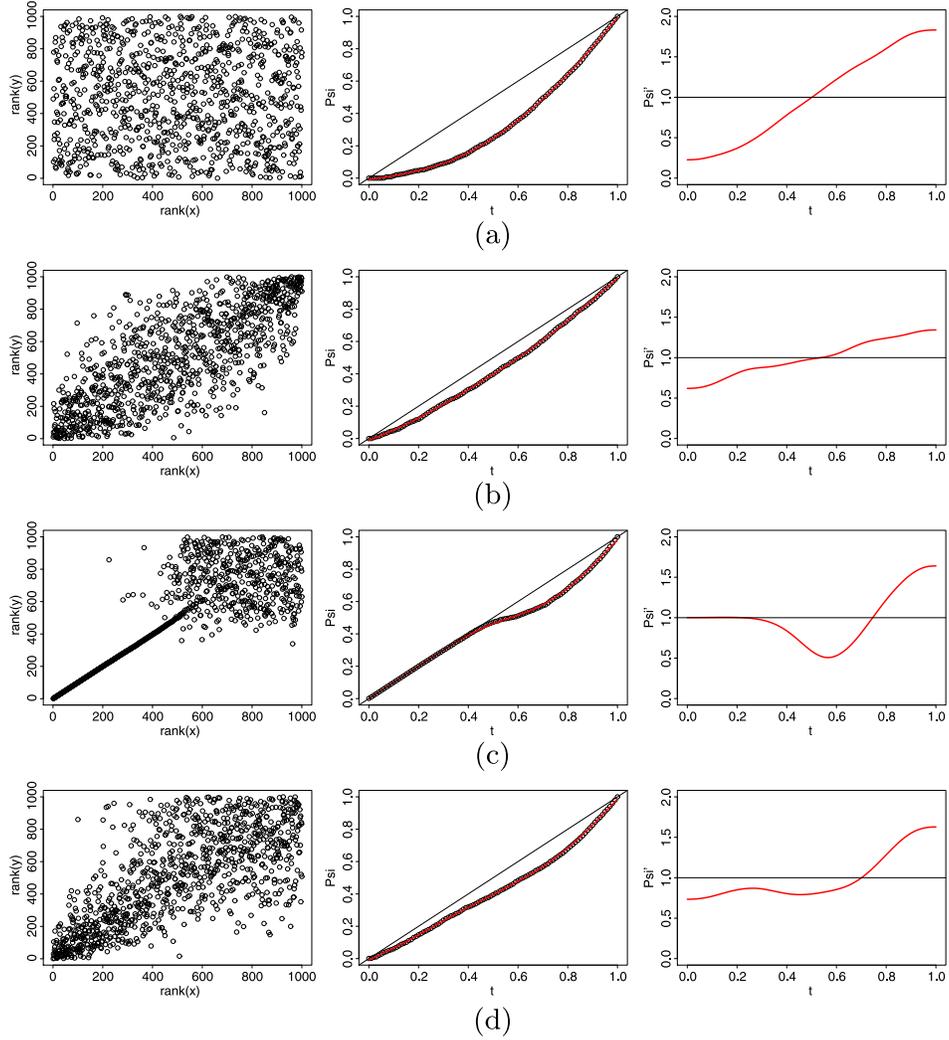}

\caption{Behavior of correspondence curves when data consists of
homogeneous and heterogeneous association. From left to right, the
three columns are the scatterplot of ranks, the curve of $\Psi$ and the curve
of $\Psi'$. \textup{(a)}  Bivariate Gaussian distribution with $\rho=0$.
\textup{(b)}~Bivariate Gaussian distribution with $\rho=0.8$. \textup{(c)}  A mixture of two
bivariate Gaussian distributions with marginals on both coordinates as
$f_{0}=N(0, 1)$ and $f_{1}=N(3, 1)$, $\rho_0=0$ and $\rho_1=1$ and
mixing proportion $\pi_1=0.5$. \textup{(d)}  A mixture of two bivariate Gaussian
distributions with marginals on both coordinates as $f_{0}=N(0, 1)$ and
$f_{1}=N(2, 1)$, $\rho_0=0$ and $\rho_1=0.8$ and mixing proportion $\pi
_1=0.5$. The curve of $\Psi_n'$ is produced by taking the derivative on
the spline that fits $\Psi_n$ with df${}={}$6.4.}\label{Fcurve}
\end{figure}

%s3.2 ###
%su3.2 #&#
\subsection{Copula mixture model}\label{sec3.2}

We first use simulation studies to examine the performance of our
approach. In particular, we aim to assess the accuracy of our
classification, to evaluate the benefit of combining information
between replicates over using only information on one replicate, and to
assess the robustness of our method to the violation of one of its
underlying model assumptions. In each simulation, we also compare the
performance with two existing methods for combining significance scores
across samples. However, as existing combination methods can be applied
to only $p$-values, we use $p$-values as the significance scores in the
comparison, though our method can be applied to arbitrary scores with
continous marginal distributions. Here we consider the scenario when
the $p$-values are not well calibrated but are reflective of the relative
strength of evidence that the signals are real, and assess the accuracy
of thresholds selected by all methods of comparison. These simulations
also provide a helpful check on the convergence of our estimation procedure.

In each simulation study, we generate a sample of $n$ pairs of signals
on two replicates. Each pair of observed signals $(Z_{i1}, Z_{i,2})$
($i=1, \ldots, n$) is a~noisy realization of a latent signal $Z_i$,
which is independently and identically generated from the following
normal mixture model:
%
%e3.1 ###
%e3.1 #&#
\begin{eqnarray}\label{Esimu}
&\displaystyle K_i  \sim  \operatorname{Bernoulli}(\pi_1),& \nonumber\\
&\displaystyle Z_i \mid K_i=k  \sim N(\mu_k, \tau_k^2), \qquad k=0, \ldots, K-1,&
\nonumber
\\[-8pt]
\\[-8pt]
&\displaystyle Z_{i,j} \mid K_i=k, Z_i  = Z_i + \epsilon_{ijk}, \qquad j=1,2,&\nonumber\\
&\displaystyle\epsilon_{ijk}  \sim N(0, \omega_k^2).&
\nonumber
\end{eqnarray}

As can easily be seen from the joint distribution of $(Z_{i,1},
Z_{i,2}) \mid K_i$,
%
%e3.2 #&#
\begin{eqnarray}
\pmatrix{Z_{i,1} \cr Z_{i,2}}\Bigm| K_i=k
\sim N \left(
\pmatrix{\mu_k \cr\mu_k},
\pmatrix{\displaystyle
\tau_k^2 + \omega_k^2 & \rho_k(\tau_k^2 + \omega_k^2) \cr\displaystyle
\rho_k(\tau_k^2 + \omega_k^2) & \tau_k^2 + \omega_k^2
}
 \right),\nonumber\\
    \eqntext{k=0, \ldots, K-1,}
\end{eqnarray}
where $\rho_k=\frac{\tau_k^2}{\tau_k^2+\omega_k^2}$;
this model is in fact a reparameterization of our mo\-del~(\ref{Estage1})
when $K=2$: $(Z_{i,1}, Z_{i,2})$ directly corresponds to
the latent Gaussian variables in (\ref{Estage1}) when setting $\mu
_0=0, \mu_1>0, \tau^2_k+\omega_k^2=1, \rho_0=0$ and $k=0, 1$. When $K >
2$, this model is convenient for simulating data from multiple
components and investigating the robustness of our method to the
violation of the assumption that the data consists of a reproducible
and an irreproducible component.
After $Z_{i, j}$ is simulated, we transform $Z_{i, j}$ to a Student's
$t$ distribution by a probability integral transformation
$F_{t_5}^{-1}(G(Z_{i, j}))$, where $F_{t_5}$ is the cdf of $t$
distribution with $\mathit{df}=5$ and $G$ is as defined in~(\ref{Estage2}). We
then obtain $p$-values from a one-sided $z$-test for $H_0\dvtx  \mu=0$ vs $H_1\dvtx
\mu>0$, and use them as the significance score $X_{i, j}$. This
procedure is equivalent to applying a $z$-test to a t distribution, thus
generating $p$-values that are not calibrated but are reflective of the
relative strength of evidence that the signals are real. It is
equivalent to letting $F_j^{-1}(\cdot)=1-\Phi(F_{t_5}^{-1}(\cdot))$ in
(\ref{Estage3}) for our model.

With $p$-values as the significance score, our method can also be viewed
as a way to combine $p$-values for ranking signals by their consensus.
The two most commonly-used methods for combining $p$-values of a set of
independent tests are Fisher's combined test [\citet{fisher1932}] and
Stouffer's $z$ method [\citet{stouffer1949}]. In Fisher's combination for
the given one-sided test, the test statistic $Q_i= -2\sum_{j=1}^{m}\log
(p_{i,j})$ for each pair of signals has the $\chi^2_{2m}$ distribution
under $H_0$, where $p_{i,j}$ is the $p$-value for the $i$th
signal on the $j$th replicate, $m$ is the number of studies
and $m=2$ here. In Stouffer's method, the test statistic $S_i= \frac
{1}{\sqrt{m}}\sum_{j=1}^m\Phi^{-1}(1-p_{i,j})$ has distribution $N(0,
1)$ under~$H_0$, where $\Phi$ is the standard normal CDF. For each pair
of signals, we compute $Q_i$ ($S_i$, resp.) and its
corresponding $p$-values $p^Q_i$ ($p^S_i$, resp.), then estimate
the corresponding false discovery rates (FDR) by computing $q$-values
[\citet{storey2003}] based on $p^Q_i$ ($p^S_i$, resp.), using R
package ``qvalue.'' FDR is estimated similarly for $p$-values on the
individual replicates.\looseness=1

For our method, we classify a call as correct (or incorrect), when a
genuine (or spurious) signal is assigned an idr value smaller than an
idr threshold. Correspondingly, for a call from individual replicates,
Fisher's method or Stouffer's method, the same classification applies,
when its corresponding $q$-value is smaller than the threshold. We
compare the discriminative power of these methods by assessing the
trade-off between the number of correct and incorrect calls made at
various thresholds.

%t1 ###
%ta1 #&#
\begin{table}
\tabcolsep=0pt
\caption{Simulation parameters and parameter estimation in the
simulation studies of 100 data sets. Each data set consists of 10\mbox{,}000
pairs of observations. The simulation parameters are estimated from a
ChIP-seq data set. In all simulations, $\mu_0=0$, $\sigma_0^2=1$ and
$\rho_0=0$. In S1--S3, $\pi_0=1-\pi_1$. S4 has a third component with
$\mu_2=0$, $\sigma_0^2=1$, $\rho_2=0.64$, $\pi_2=0.07$ and $\pi_0=1-\pi
_1-\pi_2$. The table shows the mean and the standard deviation of the
estimated parameters over the 100~data~sets~using~our~model}\label{Tsimu}
\vspace*{-3pt}
\begin{tabular*}{\textwidth}{@{\extracolsep{\fill}}lccccc@{}}
 \hline
& & $\bolds{\pi_1}$ & $\bolds{\rho_1}$ & $\bolds{\mu_1}$ & $\bolds{\sigma_1^2}$ \\
\hline
S1 & True parameter& 0.650 & 0.840 & 2.500 & 1.000 \\
& Estimated values & 0.648 (0.005) & 0.839 (0.005) & 2.524 (0.033) &
1.003 (0.024) \\[3pt]
S2 & True parameter & 0.300 & 0.400 & 2.500 & 1.000 \\
& Estimated values & 0.302 (0.004) & 0.398 (0.024) & 2.549 (0.037) &
1.048 (0.032) \\[3pt]
S3 & True parameter & 0.050 & 0.840 & 2.500 & 1.000 \\
& Estimated values & 0.047 (0.004) & 0.824 (0.026) & 2.536 (0.110) &
0.876 (0.087) \\[3pt]
S4 & True parameter & 0.650 & 0.840 & 3.000 & 1.000 \\
& Estimated values & 0.669 (0.005) & 0.850 (0.005) & 3.021 (0.031) &
1.058 (0.029) \\
\hline
\end{tabular*}
\vspace*{-5pt}
\end{table}

In an attempt to generate realistic simulations, we first estimated
parameters from a ChIP-seq data set (described in Section \ref{Sencode} using the model in Section \ref{SSinference}), then
simulated the signals on a pair of replicates using the sampling model
(\ref{Esimu}). We performed four simulations, S1, S2, S3 and S4, as
follows, with simulation parameters in Table \ref{Tsimu}:

\begin{enumerate}[S3]
\item[S1] This simulation was designed to demonstrate performance when
the data are generated from the same copula mixture model we use for
estimation. Data were simulated from the model (\ref{Esimu}) with
$K=2$, using the parameters estimated from the ChIP-seq data set
considered below. The resulting data contained $\pi_1=65\%$ signals and
$\pi_0=35\%$ noise.
\item[S2] A simulation to assess performance of our method when the
correlation between genuine signals is low. Data were simulated as in
S1 ($K=2$), except that $\rho_1=0.4$ and $\pi_1=0.3$.
\item[S3] A simulation to assess performance of our method when only a
small proportion of real but highly correlated signals are present.
Data were simulated as in S1 ($K=2$), except that $\pi_1=0.05$.
\item[S4] Here simulation parameters were chosen to illustrate a
scenario when reproducible noise is present in addition to random noise
and real signals. The goal is to assess the sensitivity of our method
to deviations from the assumption that genuine signals are reproducible
and noise is irreproducible. Data were simulated from a three-component
model (i.e., $K=3$) using (\ref{Esimu}), where $\pi_2=7\%$
reproducible noise is added as the third component with\vspace*{1pt} $\rho_2=0.64$,
$\mu_2=0$ and $\sigma_2^2=1$, and the parameters for signals and random
noise are as in S1, except $\pi_0=28\%$ and $\mu_1=3$.
\end{enumerate}

For each parameter set, we simulated 100 data sets, each of which
consists of two replicates with 10\mbox{,}000 signals on each replicate. In
each simulation, we ran the estimation procedure from 10 random
initializations, and stopped the procedure when the increment of
log-likelihood is $< $0.01 in an iteration or the number of iterations
exceeds 100. All the simulations converge, when starting points are
close to the true parameters. The results that converge to the highest
likelihood are reported.\vadjust{\eject}

%s3.2.1 ###
%su3.2.1 #&#
\subsubsection{Parameter estimation and calibration of IDR}\label{sec3.2.1}

In S1--S3, the parameters estimated from our models are close to the
true parameters (Table~\ref{Tsimu}). The only exception is that $\sigma
_1$ was underestimated when the proportion of true signals is small,
$\pi_1=0.05$, a case hard to distinguish from that of a single component.

%f4 ###
%fi4 #&#
\begin{figure}

\includegraphics{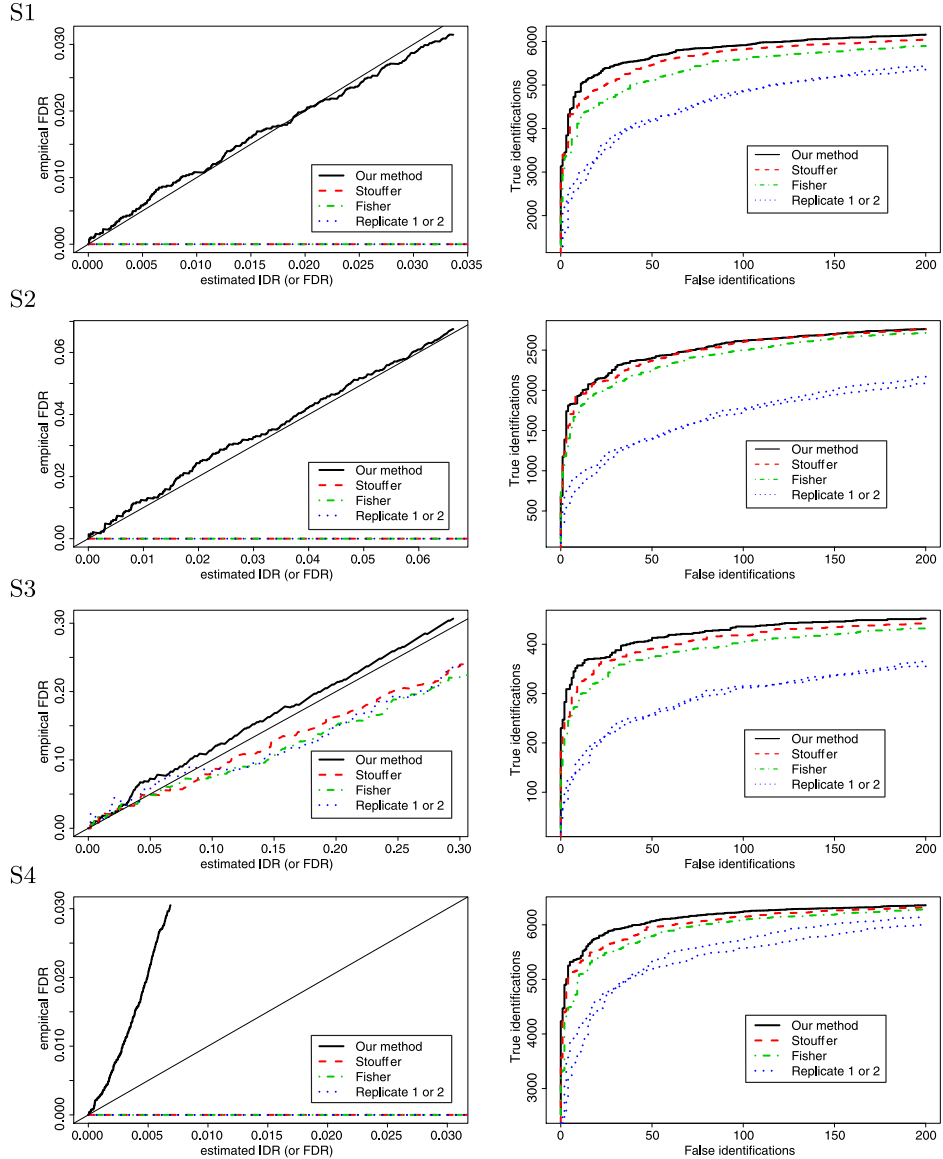}

\caption{Calibration of IDR (Left) and comparison of discriminative
power (Right) in simulation studies. Left: Estimated error rate
($x$-axis: IDR for our method and FDR for other methods) is compared with
the actual frequency of false identifications ($y$-axis). Right: The
number of correct and incorrect calls made at various thresholds in
simulation studies. Incorrect calls: The number of spurious signals
assigned idr values smaller than the thresholds (our method) or with
$q$-values smaller than the cutoffs
(other methods). Correct calls: The number of genuine signals assigned
idr values smaller than the thresholds (our method) or with $q$-values
smaller than the cutoffs
(other methods).}\label{Fsimu1}
\end{figure}

The irreproducible discovery rate as a guide for the selection of the
signals needs to be well calibrated.
To check the calibration of thresholds, we compare the actual frequency
of false calls, that is, empirical FDR, with the estimated IDR for our
method and with the $q$-values for other methods (Figure \ref{Fsimu1},
left column).

As shown in Figure \ref{Fsimu1}, the original significance scores and
other combination methods are overly conservative in their estimated
FDR in all simulations, whereas our method is reasonably well
calibrated in S1, S2 and S3.
When reproducible noise is present (S4), our method slightly
overestimates the proportion and the correlation of the real signals,
and underestimates the empirical FDR (Figure \ref{Fsimu1}-S4). This
reflects that the data contains some artifacts that receive
reproducible high scores on their original measures and consequently
receive relatively low idr values. These artifacts are difficult to
distinguish from genuine signals. We will compare the discriminative
power of all four methods in the next section.

%s3.2.2 ###
%su3.2.2 #&#
\subsubsection{Comparison of discriminative power}\label{sec3.2.2}\label{SSdisc}

To assess the benefit of combining information on replicates and
compare with existing methods of combining $p$-values, we compared our
method with the $p$-values on individual replicates, Fisher's method and
Stouffer's method, by assessing the trade-off between the numbers of
correct and incorrect calls made at various thresholds. As a small
number of false calls is desired in practice, the comparison focuses on
the performance in this region.

In all simulations, our method consistently identifies more true
signals than the original significance score and the two $p$-value
combination\vadjust{\goodbreak} methods, at a given number of false calls in the studied
region. Even when reproducible artifacts are present (S4) or only a
small proportion of genuine signal is present (S3), our method still
outperforms all methods compared here.

%s4 ###
%se4 #&#
\section{Applications on real data}\label{sec4}\label{Sencode}

%s4.1 ###
%su4.1 #&#
\subsection{Comparing the reproducibility of multiple peak callers for
ChIP-seq experiments}\label{sec4.1}\label{SScomparison}

We now consider an application arising from a collaborative project
with the ENCODE consortium [\citet{encode2004}]. This project has three
primary goals: comparing the reproducibility of multiple algorithms for
identifying protein-binding regions in ChIP-seq data (described below),
selecting binding regions using a uniform criterion for data from
different sources (e.g., labs), and identifying experiment results in
poor quality.

We now state the background of ChIP-Seq data in more detail and refer
to \citet{park2009} for a recent review. A ChIP-seq experiment is a
high-throughput assay to study protein binding sites on DNA sequences
in a genome. In a~typical ChIP-seq experiment, DNA regions that are
specifically bound by the protein of interest are first enriched by
immunoprecipitation, then the enriched DNA regions are sequenced by
high-throughput sequencing, which generates a genome-wide scan of tag
counts that correspond to the level of enrichment at each region.
The relative significance of the regions are determined by a
computational algorithm (usually referred to as a peak caller), largely
according to local tag counts, based on either heuristics or some
probabilistic models. The regions whose significance are above some
prespecified threshold then are identified. To date, more than a dozen
of the peak callers have been published. Some common measures of
significance are fold enrichment, $p$-value or $q$-value [\citet{storey2003}].

Though these scores may reflect the relative strength of evidence for
putative binding regions to be real, determination of a proper
threshold is not straightforward, especially for heuristic-based
scores, where arbitrary judgment often has to be involved. In fact,
this difficulty could also exist for probabilistic-based scores if the
underlying probabilistic models are inadequate to capture the
complexity of the data. Because tuning parameters for each data set are
usually infeasible due to lack of ground truth, default thresholds are
often used in practice, though they may not be the optimal choices for
the data to be analyzed. Ideally, an objective performance assessment
should reflect the behavior of peak callers instead of the effect of thresholds.

Here we use the binding regions identified at untuned thresholds in
a~CTCF ChIP-seq experiment (described below) to illustrate how our method
is used for assessing and comparing the reproducibility of peak callers
when tuning thresholds are unavailable, for setting a
reproducibility-based threshold that is appliable to both heuristic and
probabilistic-based significance scores, and for identifying results
with low reproducibility. A detailed analysis on a~comprehensive set of
ENCODE data will appear elsewhere.

%s4.1.1 ###
%su4.1.1 #&#
\subsubsection{Description of the data}\label{sec4.1.1}\label{sssdata}

In this comparison the ChIP-seq experiments of a transcription factor
CTCF from two biological replicates were generated from the Bernstein
Laboratory at the Broad Institute on human K526 cells.
Peaks were identified in biology labs, using nine commonly used and
publicly available peak callers, namely, Peakseq [\citet{peakseq}], MACS
[\citet{macs}], SPP [\citet{spp}], Fseq [\citet{fseq}], Hotspot [\citet{hotspot}], Erange [\citet{erange}], Cisgenome [\citet{cisgenome}], Quest
[\citet{quest}] and SISSRS [\citet{sissrs}], using their default
significance measures and default parameter settings with either
default thresholds (all peak callers except Hotspot) or more relaxed
thresholds (Hotspot). Among them, Peakseq and SPP use $q$-value, MACS,
Hotspot and SISSRS use $p$-value, and the rest use fold enrichment, as
their significance measures. Only the outputs from peak callers were
available for our analysis.

The peaks generated from different algorithms have substantially
different peak widths.
SPP and SISSRS generate peaks with fixed width of 100~bp and 40~bp,
respectively; all other algorithms generate peaks with varying peak
width (median${}={}$130--760~bp).
Because wider peaks are more likely to hit true binding sites by
chance than shorter peaks, we normalized peak width by truncating the
peaks wider than 40~bp down to intervals of 40~bp centered at the
reported summits of peaks, so that reproducibility is compared on the
same basis. The choice of 40~bp was made because the peak caller with
the narrowest average peak width in our comparison reports peaks with a
fixed width of 40~bp.
Prior to applying our method, peaks on different replicates are paired
as identifying the same binding region, if their coverage regions
overlap (i.e., overlap $\geq$ 1~bp).
Because peaks without matches do not have replicate measurements and
are apparently irreproducible, here we elected to assess
reproducibility of paired peaks in our analysis. Around 23--78\% of
peaks are retained for this analysis.

%f5 ###
%fi5 #&#
\begin{figure}[b]

\includegraphics{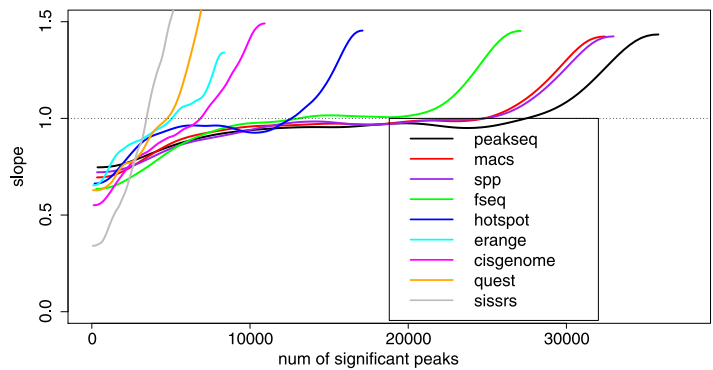}

\caption{The change of correspondence ($\Psi_n'$) along the decreasing
order of significance, plotted for 9 peak callers on a CTCF ChIP-seq
experiment from ENCODE. $X$-axis: The rank list of peaks identified on a
replicate. $Y$-axis: $\Psi_n'$.}\label{Fprofile-nonmissing}
\end{figure}

%s4.2 ###
%su4.2 #&#
\subsection{Results}\label{sec4.2}\label{SSresults}

%s4.2.1 ###
%su4.2.1 #&#
\subsubsection{Correspondence profiles}\label{sec4.2.1}

Figure \ref{Fprofile-nonmissing} shows the correspondence profiles for
the nine peak callers. By referring to the prototypical plots in Figure
\ref{Fcurve}, five peak callers (Peakseq, MACS, SPP, Fseq and Hotspot)
show the characteristic transition from strong association to near
independence [Figure \ref{Fprofile-nonmissing}(b)]. As described in
Section \ref{SSprofile}, when heterogeneity of association is present,
a high reproducibility translates to late occurrence of the transition
to a segment with a positive slope. According to how much down the rank
list the transition is observed, the three peak callers that show the
highest reproducibility on this data set are Peakseq, MACS and SPP
(Figure \ref{Fprofile-nonmissing}). For the other four peak callers
(Erange, Cisgenome, Quest and SISSRS),
the curves display a less clear transition and report substantially
fewer (reproducible) peaks. This indicates that the default thresholds
for these peak callers are likely to be too stringent to reach the
breakdown of consistency, and that the reported peaks have relatively
low reproducibility across replicates. This conclusion was confirmed
later by biological verification (see Section \ref{SSbiology}).

%s4.2.2 ###
%su4.2.2 #&#
\subsubsection{Inference from the copula mixture model}\label{sec4.2.2}

We applied the copula mixture model to the peaks identified on the
replicates for each peak caller. As data may consist of only one group
with homogeneous association, we also estimated the fit using a
one-component model that corresponds to setting $\pi_1=1, \mu_1=0$ and
$\sigma_1^2=1$ in (\ref{Emixture}). We then tested for the smallest
number of\vspace*{-2pt} components compatible with the data, using a likelihood ratio
test statistic\vspace*{1pt} ($\lambda=\frac{L_2}{L_1}$), where $L_2$ and $L_1$ are
the likelihood of two-component and one-component models, respectively.
With mixture models, it is well known that the regularity conditions do
not hold for $2\log(\lambda)$ to have its usual asymptotic Chi-square
null distribution. We therefore used a parametric bootstrap procedure
to obtain appropriate $p$-values [\citet{mclachlan1987}]. In our procedure,
100 bootstrap samples were sampled from the null distribution under the
one component hypothesis using the parametric bootstrap, where the
parameter estimate was obtained by maximizing the pseudo-likelihood of
the data under the null hypothesis of the one-component model. Then
$p$-values were obtained by referring to the distribution of the
likelihood ratio computed from the bootstrap samples. Table \ref{Tpara-real} summarizes the parameter estimation from both models and
the bootstrap results.

Based on the likelihood ratio test, it seems that the one-component
model fits the results from SISSRS, Quest and Cisgenome better, and the
two-component model fits the results from other peak callers. This is
consistent with the pattern of transition in the correspondence
profiles (Figure \ref{Fprofile-nonmissing}).

To select binding sites, we rank putative peaks by the values of local
idr and compute the irreproducible discovery rate (IDR) for peaks
selected at various local idr cutoffs using (\ref{EIDR}), as described
in Section \ref{SSidr}. We illustrate the IDR as a function of the
numbers of top peaks (ranked by local idr) for all peak callers in
Figure \ref{Fposterior}. For a given IDR level, one can determine the
number of peaks to be called from this plot, regardless of what type of
scores are used to measure the significance of peaks.
For example, at $5\%$ IDR, the top 27\mbox{,}500 peaks with the smallest local
idr can be called using MACS. Using the same reproducibility criterion,
peaks can be selected for other peak callers similarly.

%
%t2 ###
%ta2 #&#
\begin{table}
\tabcolsep=0pt
\caption{Parameters estimated from the copula mixture model and the
one-component model, and model selection for determining the number of
components. $(\pi_1, \rho_1, \mu_1, \sigma_1)$ are parameters estimated
from the copula mixture model; $\rho$ is estimated from the
single-component model. The number of components is selected using a
likelihood ratio test and the $p$-value of the test statistics is
determined using a parametric bootstrap approach based on 100 bootstrap
samples}\label{Tpara-real}
\vspace*{-3pt}
\begin{tabular*}{\textwidth}{@{\extracolsep{\fill}}lccccccccc@{}}
\hline
&\textbf{Peakseq}&\textbf{MACS}&\textbf{SPP}&\textbf{Fseq}&\textbf{Hotspot}&\textbf{Cisgenome}&\textbf{Erange}
&\textbf{Quest}&\textbf{Sissrs}\\
\hline
$\pi_1$ & 0.69 & 0.84 & 0.77 & 0.74 & 0.69 & 0.85 & 0.72 & 0.72 & 1\hphantom{.00}\\
$\rho_1$ & 0.89 & 0.89 & 0.88 & 0.82 & 0.88 & 0.65 & 0.81 & 0.67 &
0.24\\
$\mu_1$ & 2.27 & 2.07 & 2.28 & 2.12 & 1.62 & 2.05 & 2.04 & 2.01 & 7.27
\\
$\sigma_1$ & 0.87 & 1.34 & 1.05 & 0.86 & 0.64 & 1.35 & 0.90 & 1.39 &
0.03 \\[3pt]
$\rho$ & 0.87 & 0.87 & 0.86 & 0.83 & 0.78 & 0.66 & 0.80 & 0.66 & 0.23
\\[3pt]
$p $-value & 0\hphantom{.00} & 0\hphantom{.00} & 0\hphantom{.00} & 0\hphantom{.00} & 0\hphantom{.00} & 1\hphantom{.00} & 0\hphantom{.00}
 & 1\hphantom{.00} & 1\hphantom{.00} \\ \hline
\end{tabular*}
\vspace*{-3pt}
\end{table}
%

%f6 ###
%fi6 #&#
\begin{figure}[b]
\vspace*{-3pt}
\includegraphics{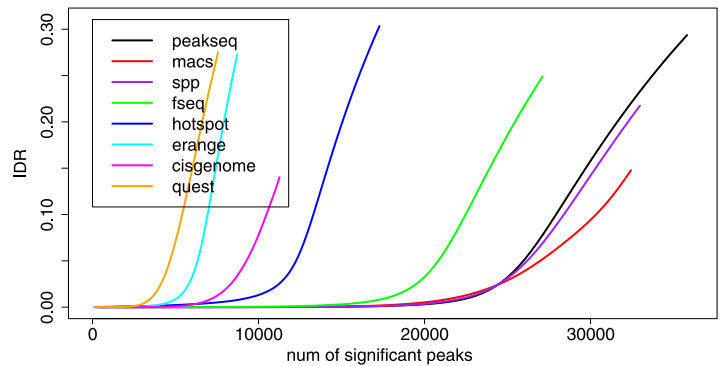}%
\vspace*{-3pt}
\caption{Irreproducible discovery rate (IDR) at different numbers of selected
peaks, plotted at various idr cutoffs for eight peak callers on a CTCF
Chip-seq experiment from ENCODE. Peaks are selected using local idr.
$X$-axis: The rank list of peaks,
ranked by local idr, $Y$-axis:
Irreproducible discovery rate (IDR). SISSRS is not shown because its
results are highly inconsistent and all peaks are grouped into a low
correlation group.}\label{Fposterior}
\end{figure}

We also compare the overall reproducibility of different peak callers
using Figure \ref{Fposterior}. For example, while Peakseq, MACS and
SPP on average have about $3\%$ irreproducible peaks when selecting the
top 25\mbox{,}000 peaks, most of the other peak callers have already reached a
much higher IDR before identifying the top 10\mbox{,}000 peaks.
According to the number of peaks identified before reaching $5\%$ IDR,
the three most reproducible peak callers on this data set are Peakseq,
MACS and SPP, then followed by Fseq, then others. This result is
consistent with the graphical comparison based on the correspondence
profile (Figure \ref{Fprofile-nonmissing}).\vadjust{\eject}

%s4.2.3 ###
%su4.2.3 #&#
\subsubsection{Evaluating the biological relevance of the
reproducibility assessment}\label{sec4.2.3}\label{SSbiology}

To evaluate the biological relevance of our reproducibility assessment,
we check the accuracy of peak identifications using external biological
information.
Because a complete list of true binding regions is not known for the
examined data set, the accuracy of peak identifications is assessed
using high-confidence binding motifs computationally predicted using
sequence information [\citet{kheradpour2007}], which is a commonly used
device in this setting [e.g., \citet{macs}; \citet{spp}, among many others].
Though high-confidence motifs are not required to be bound and true
binding sites are not required to exhibit a motif signature, the
high-confidence motif instances are assumed, standard in this context,
to contain a representative subset of true binding regions and are
expected to have a relatively high occurrence in high-scored ChIP-seq
peaks [\citet{spp}].
We selected ChIP-seq peaks reported by each peak caller at various IDR
thresholds, and examined the number of high-confidence motifs (FDR
$\leq0.1$ at the PWM threshold of $p$-value${}=\frac{1}{4^{10}}$) that
coincide with the reported ChIP-seq peaks\vspace*{2pt} (defined as overlap${}\geq
1$~bp) (Figure \ref{Faccuracy}).

%f7 ###
%fi7 #&#
\begin{figure}[b]

\includegraphics{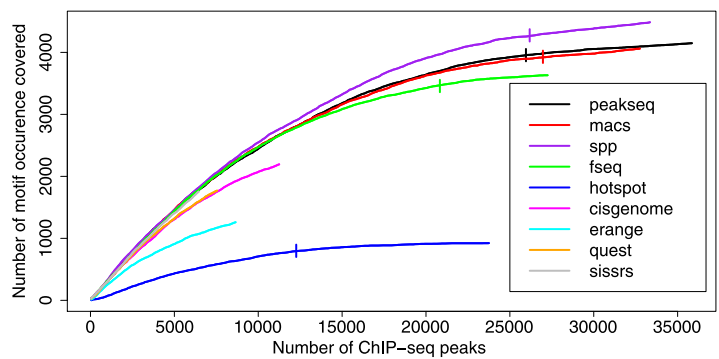}

\caption{The coverage of high-confidence CTCF motif at different
numbers of selected ChIP-seq peaks, plotted at various idr cutoffs for
nine peak callers on a CTCF Chip-seq experiment from ENCODE. The bars
on the curves of Peakseq, MACS, SPP, Fseq and Hotspot show  the
number of peaks selected  at IDR${}={}$0.05. No selection is made for the rest
of the peak callers because model selection favors the one-component
model for peaks identified by these callers.}\label{Faccuracy}
\end{figure}

For the peak callers whose reported peaks fit the two-component model
(i.e., Peakseq, MACS, SPP, Fseq and Hotspot), we marked the number of
ChIP-seq peaks selected at IDR${}={}$5\%. For these algorithms, the motif
occurrences first increase with the increase of reported ChIP-seq
peaks, then plateau before reaching the default thresholds (Figure \ref{Faccuracy}). The mark of $5\%$ IDR approximately corresponds to the
occurrences of the plateau, with few additional motif occurrences if
more ChIP-seq peaks are called.

On the other hand, for the peak callers (Erange, Cisgenome, Quest and
SISSRS) whose peaks fit the one-component model, the motif occurrence
still shows an increasing trend at default thresholds. This confirms
the observation from correspondence curves (Figure \ref{Fprofile-nonmissing}) that the default thresholds for these peak
callers are likely to be overly stringent for this data set.

Overall, the results of this analysis agree with the assessment from
our reproducibility comparison: the three peak calling results with the
highest reproducibility (SPP, Peakseq and MACS) in Figure \ref{Fposterior} show the highest rates of motif occurrence among all
algorithms of comparison; the ones that are reported to be less
reproducible do show lower rates of motif occurrence. This illustrates
the potential of our method as a quality measure.

%s5 ###
%se5 #&#
\section{Discussion}\label{sec5}\label{Sdiscussion}

We have presented a new statistical method for measuring the
reproducibility of results in high-throughput experiments and setting
selection thresholds using a reproducibility criterion.
Using simulated and real data, we have illustrated the potential of our
method for providing reproducibility assessment that is not confounded
with prespecified threshold choices, determining biologically relevant
thresholds, improving the accuracy of signal identification, and
identifying suboptimal results.

As no assumption is made on the scale of the scores,
the proposed method is applicable for any scoring system that produces
continuous ranking to reflect the relative ordering of the signals.
It provides a principled way to select signals that are scored on
heuristic measures, and complements the thresholds determined on
individual replicates. Moreover, because consistency between replicates
is an internal standard that is independent of the scoring schemes and
comparable across data sets, the proposed reproducibility criterion is
suited for setting uniform standards for selecting signals for data
from multiple sources, such as consortium studies. Because our measure
of consistency is not confounded by platform-dependent thresholds,
inter-platform consistency can be assessed easily.

Of course, reproducibility is only a necessary but not a sufficient
condition to accuracy. If replicates are generated in the presence of a
systematic bias that introduces false association, the threshold
derived from this procedure may underestimate the empirical false
discovery rate. Though the thresholds determined by our method show
reasonable biological relevance in the data examined here and many
other ENCODE ChIP-seq experiments (to appear in another manuscript), we
emphasize that some cares are necessary to ensure that the replicates
maintain the level of independence that they should.

We also note that the reproducibility of outputs from a data-analytical
method (e.g., a peak caller) on replicate samples reflects the combined
properties of the method and the samples. As the behavior of the
data-analytical method may vary across different samples, the
reproducibility assessment in our example should be interpreted as
being specific to the studied data set, instead of a general
conclusion. A detailed comparison of the performance of peak callers
has been evaluated on a comprehensive set of data and will appear elsewhere.

The algorithm to implement the estimation strategy outlined in Section~\ref{sec2.2.3}
is provided as supplemental material to this article. An R
package is downloadable at the following website:
\href{http://cran.r-project.org/web/packages/idr/index.html}{http://cran.r-project.org/web/}\break
\href{http://cran.r-project.org/web/packages/idr/index.html}{packages/idr/index.html}.

\begin{supplement}%[id=suppA]
\stitle{Supplementary materials for Measuring reproducibility of
high-through\-put experiments}
\slink[doi]{10.1214/11-AOAS466SUPP} %[doi,text={...}] - jei reikia
%suskaldyti doi
\slink[url]{http://lib.stat.cmu.edu/aoas/466/supplement.pdf}
\sdatatype{.pdf}
\sdescription{This supplement consists of four parts. Part~1 describes
the algorithm for estimating parameters in our
copula mixture model. Part~2 provides a theoretical justification for
the efficiency of our estimator for the
proposed copula mixture model when n is large. Part~3 derives the
properties of the correspondence curves in
Section~\ref{sec2.1.1}. Part~4 provides an extension of our model to the case
with multiple replicates.}
\end{supplement}

\section*{Acknowledgments}
We thank Ewan Birney, Ian Dunham, Anshul Kundaje and Joel Rozowsky for
helpful discussions,
Pouya Kheradpour and Manolis Kellis for providing CTCF motif prediction
and the ENCODE element group for
generating the peak calling results.

% imsref loaded by smiklovaite, 2011-04-29 08:53:34
% imsref loaded by smiklovaite, 2011-04-29 09:59:02

\printaddresses


\begin{thebibliography}{42}
% BibTex style file: ims.bst, 2010-03-23
% Default style options (sort=0,type=number).
% Used options (sort=1,type=nameyear).

%b1 ###
%bi1 #&#
\bibitem[\protect\citeauthoryear{Benjamini and Hochberg}{1995}]{benjamini1995}
\begin{barticle}[mr]
\bauthor{\bsnm{Benjamini},~\bfnm{Yoav}\binits{Y.}} \AND
  \bauthor{\bsnm{Hochberg},~\bfnm{Yosef}\binits{Y.}}
(\byear{1995}).
\btitle{Controlling the false discovery rate: A practical and powerful approach
  to multiple testing}.
\bjournal{J. Roy. Statist. Soc. Ser. B}
\bvolume{57}
\bpages{289--300}.
\bid{issn={0035-9246}, mr={1325392}}
\end{barticle}
\endbibitem

%b2 ###
%bi2 #&#
\bibitem[\protect\citeauthoryear{Blest}{2000}]{blest2000}
\begin{barticle}[mr]
\bauthor{\bsnm{Blest},~\bfnm{David~C.}\binits{D.~C.}}
(\byear{2000}).
\btitle{Rank correlation---an alternative measure}.
\bjournal{Aust. N. Z. J. Stat.}
\bvolume{42}
\bpages{101--111}.
\bid{doi={10.1111/1467-842X.00110}, issn={1369-1473}, mr={1747465}}
\end{barticle}
\endbibitem

%b3 ###
%bi3 #&#
\bibitem[\protect\citeauthoryear{Boulesteix and Slawski}{2009}]{boulesteix2009}
\begin{barticle}[auto:STB|2011-03-03|12:04:44]
\bauthor{\bsnm{Boulesteix},~\bfnm{A.~L.}\binits{A.~L.}} \AND
  \bauthor{\bsnm{Slawski},~\bfnm{M.}\binits{M.}}
(\byear{2009}).
\btitle{Stability and aggregation of ranked gene lists}.
\bjournal{Briefings in Bioinformatics}
\bvolume{10}
\bpages{556--568}.
\end{barticle}
\endbibitem

%b4 ###
%bi4 #&#
\bibitem[\protect\citeauthoryear{Boyle et~al.}{2008}]{fseq}
\begin{barticle}[auto:STB|2011-03-03|12:04:44]
\bauthor{\bsnm{Boyle},~\bfnm{A.~P.}\binits{A.~P.}},
  \bauthor{\bsnm{Guinney},~\bfnm{J.}\binits{J.}},
  \bauthor{\bsnm{Crawford},~\bfnm{G.~E.}\binits{G.~E.}} \AND
  \bauthor{\bsnm{Furey},~\bfnm{T.~S.}\binits{T.~S.}}
(\byear{2008}).
\btitle{F-Seq: A feature density estimator for high-throughput sequence tags}.
\bjournal{Bioinformatics}
\bvolume{24}
\bpages{2537--2538}.
\end{barticle}
\endbibitem

%b5 ###
%bi5 #&#
\bibitem[\protect\citeauthoryear{da~Costa and Soares}{2005}]{dacosta2005}
\begin{barticle}[mr]
\bauthor{\bparticle{da} \bsnm{Costa},~\bfnm{Joaquim~Pinto}\binits{J.~P.}} \AND
  \bauthor{\bsnm{Soares},~\bfnm{Carlos}\binits{C.}}
(\byear{2005}).
\btitle{A weighted rank measure of correlation}.
\bjournal{Aust. N. Z. J. Stat.}
\bvolume{47}
\bpages{515--529}.
\bid{doi={10.1111/j.1467-842X.2005.00413.x}, issn={1369-1473}, mr={2235420}}
\end{barticle}
\endbibitem

%b6 ###
%bi6 #&#
\bibitem[\protect\citeauthoryear{Deheuvels}{1979}]{deheuvels1979}
\begin{barticle}[mr]
\bauthor{\bsnm{Deheuvels},~\bfnm{Paul}\binits{P.}}
(\byear{1979}).
\btitle{La fonction de d\'ependance empirique et ses propri\'et\'es. {U}n test
  non param\'etrique d'ind\'ependance}.
\bjournal{Acad. Roy. Belg. Bull. Cl. Sci. (5)}
\bvolume{65}
\bpages{274--292}.
\bid{issn={0001-4141}, mr={0573609}}
\end{barticle}
\endbibitem

%b7 ###
%bi7 #&#
\bibitem[\protect\citeauthoryear{Dempster, Laird and
  Rubin}{1977}]{dempster1977}
\begin{barticle}[mr]
\bauthor{\bsnm{Dempster},~\bfnm{A.~P.}\binits{A.~P.}},
  \bauthor{\bsnm{Laird},~\bfnm{N.~M.}\binits{N.~M.}} \AND
  \bauthor{\bsnm{Rubin},~\bfnm{D.~B.}\binits{D.~B.}}
(\byear{1977}).
\btitle{Maximum likelihood from incomplete data via the {EM} algorithm}.
\bjournal{J. Roy. Statist. Soc. Ser. B}
\bvolume{39}
\bpages{1--38}.
\bid{issn={0035-9246}, mr={0501537}}
\bptnote{check related}%
\end{barticle}
\endbibitem

%b8 ###
%bi8 #&#
\bibitem[\protect\citeauthoryear{Efron}{2004}]{efron2004a}
\begin{bmisc}[auto:STB|2011-03-03|12:04:44]
\bauthor{\bsnm{Efron},~\bfnm{B.}\binits{B.}}
(\byear{2004}).
\bhowpublished{Local false discovery rate. Technical report, Dept.   Statistics,
  Stanford Univ.}
\end{bmisc}
\endbibitem

%b9 ###
%bi9 #&#
\bibitem[\protect\citeauthoryear{ENCODE Project Consortium}{2004}]{encode2004}
\begin{bmisc}[pbm]
\borganization{ENCODE Project Consortium}
(\byear{2004}).
\bhowpublished{The ENCODE (ENCyclopedia Of DNA Elements) Project.
\textit{Science}
\textbf{306}
 636--640}.
\bid{doi={10.1126/science.1105136}, issn={1095-9203}, pii={306/5696/636},
  pmid={15499007}}
\end{bmisc}
\endbibitem

%b10 ###
%bi10 #&#
\bibitem[\protect\citeauthoryear{Fisher}{1925}]{fisher1932}
\begin{bbook}[auto:STB|2011-03-03|12:04:44]
\bauthor{\bsnm{Fisher},~\bfnm{R.~A.}\binits{R.~A.}}
(\byear{1925}).
\btitle{Statistical Methods for Research Workers},
\bedition{1st} ed.
\bpublisher{Oliver \& Boyd},
\baddress{Edinburgh}.
\end{bbook}
\endbibitem

%b11 ###
%bi11 #&#
\bibitem[\protect\citeauthoryear{Fisher and Switzer}{1985}]{fisher1985}
\begin{barticle}[mr]
\bauthor{\bsnm{Fisher},~\bfnm{N.~I.}\binits{N.~I.}} \AND
  \bauthor{\bsnm{Switzer},~\bfnm{P.}\binits{P.}}
(\byear{1985}).
\btitle{Chi-plots for assessing dependence}.
\bjournal{Biometrika}
\bvolume{72}
\bpages{253--265}.
\bid{doi={10.1093/biomet/72.2.253}, issn={0006-3444}, mr={0801767}}
\end{barticle}
\endbibitem

%b12 ###
%bi12 #&#
\bibitem[\protect\citeauthoryear{Fisher and Switzer}{2001}]{fisher2001}
\begin{barticle}[mr]
\bauthor{\bsnm{Fisher},~\bfnm{N.~I.}\binits{N.~I.}} \AND
  \bauthor{\bsnm{Switzer},~\bfnm{P.}\binits{P.}}
(\byear{2001}).
\btitle{Graphical assessment of dependence: Is a picture worth 100 tests?}
\bjournal{Amer. Statist.}
\bvolume{55}
\bpages{233--239}.
\bid{doi={10.1198/000313001317098248}, issn={0003-1305}, mr={1963399}}
\end{barticle}
\endbibitem

%b13 ###
%bi13 #&#
\bibitem[\protect\citeauthoryear{Genest and Boies}{2003}]{genest2003}
\begin{barticle}[mr]
\bauthor{\bsnm{Genest},~\bfnm{Christian}\binits{C.}} \AND
  \bauthor{\bsnm{Boies},~\bfnm{Jean-Claude}\binits{J.-C.}}
(\byear{2003}).
\btitle{Detecting dependence with {K}endall plots}.
\bjournal{Amer. Statist.}
\bvolume{57}
\bpages{275--284}.
\bid{doi={10.1198/0003130032431}, issn={0003-1305}, mr={2016261}}
\end{barticle}
\endbibitem

%b14 ###
%bi14 #&#
\bibitem[\protect\citeauthoryear{Genest, Ghoudi and Rivest}{1995}]{genest1995}
\begin{barticle}[mr]
\bauthor{\bsnm{Genest},~\bfnm{C.}\binits{C.}},
  \bauthor{\bsnm{Ghoudi},~\bfnm{K.}\binits{K.}} \AND
  \bauthor{\bsnm{Rivest},~\bfnm{L.~P.}\binits{L.~P.}}
(\byear{1995}).
\btitle{A semiparametric estimation procedure of dependence parameters in
  multivariate families of distributions}.
\bjournal{Biometrika}
\bvolume{82}
\bpages{543--552}.
\bid{doi={10.1093/biomet/82.3.543}, issn={0006-3444}, mr={1366280}}
\end{barticle}
\endbibitem

%b15 ###
%bi15 #&#
\bibitem[\protect\citeauthoryear{Genest and Plante}{2003}]{genest2003b}
\begin{barticle}[mr]
\bauthor{\bsnm{Genest},~\bfnm{Christian}\binits{C.}} \AND
  \bauthor{\bsnm{Plante},~\bfnm{Jean-Fran{\c{c}}ois}\binits{J.-F.}}
(\byear{2003}).
\btitle{On {B}lest's measure of rank correlation}.
\bjournal{Canad. J. Statist.}
\bvolume{31}
\bpages{35--52}.
\bid{doi={10.2307/3315902}, issn={0319-5724}, mr={1985503}}
\end{barticle}
\endbibitem

%b16 ###
%bi16 #&#
\bibitem[\protect\citeauthoryear{Genovese and Wasserman}{2002}]{genovese2002}
\begin{barticle}[mr]
\bauthor{\bsnm{Genovese},~\bfnm{Christopher}\binits{C.}} \AND
  \bauthor{\bsnm{Wasserman},~\bfnm{Larry}\binits{L.}}
(\byear{2002}).
\btitle{Operating characteristics and extensions of the false discovery rate
  procedure}.
\bjournal{J. R. Stat. Soc. Ser. B Stat. Methodol.}
\bvolume{64}
\bpages{499--517}.
\bid{doi={10.1111/1467-9868.00347}, issn={1369-7412}, mr={1924303}}
\end{barticle}
\endbibitem

%b17 ###
%bi17 #&#
\bibitem[\protect\citeauthoryear{Hu}{2006}]{hu2006}
\begin{barticle}[auto:STB|2011-03-03|12:04:44]
\bauthor{\bsnm{Hu},~\bfnm{L.}\binits{L.}}
(\byear{2006}).
\btitle{Dependence patterns across financial markets: A mixed copula approach}.
\bjournal{Applied Financial Economics}
\bvolume{16}
\bpages{717--729}.
\end{barticle}
\endbibitem

%b18 ###
%bi18 #&#
\bibitem[\protect\citeauthoryear{Ji et~al.}{2008}]{cisgenome}
\begin{barticle}[auto:STB|2011-03-03|12:04:44]
\bauthor{\bsnm{Ji},~\bfnm{H.}\binits{H.}},
  \bauthor{\bsnm{Jiang},~\bfnm{H.}\binits{H.}},
  \bauthor{\bsnm{Ma},~\bfnm{W.}\binits{W.}},
  \bauthor{\bsnm{Johnson},~\bfnm{D.~S.}\binits{D.~S.}},
  \bauthor{\bsnm{Myers},~\bfnm{R.~M.}\binits{R.~M.}} \AND
  \bauthor{\bsnm{Wong},~\bfnm{W.~H.}\binits{W.~H.}}
(\byear{2008}).
\btitle{An integrated software system for analyzing ChIP-chip and ChIP-seq
  data}.
\bjournal{Nature Biotechnology}
\bvolume{26}
\bpages{1293--1300}.
\end{barticle}
\endbibitem

%b19 ###
%bi19 #&#
\bibitem[\protect\citeauthoryear{Joe}{1997}]{joe1997}
\begin{bbook}[mr]
\bauthor{\bsnm{Joe},~\bfnm{Harry}\binits{H.}}
(\byear{1997}).
\btitle{Multivariate Models and Dependence Concepts}.
\bseries{Monogr. Statist. Appl. Probab.}
\bvolume{73}.
\bpublisher{Chapman \& Hall}, \baddress{London}.
\bid{mr={1462613}}
\end{bbook}
\endbibitem

%b20 ###
%bi20 #&#
\bibitem[\protect\citeauthoryear{Jothi et~al.}{2008}]{sissrs}
\begin{barticle}[auto:STB|2011-03-03|12:04:44]
\bauthor{\bsnm{Jothi},~\bfnm{R.}\binits{R.}},
  \bauthor{\bsnm{Cuddapah},~\bfnm{S.}\binits{S.}},
  \bauthor{\bsnm{Barski},~\bfnm{A.}\binits{A.}},
  \bauthor{\bsnm{Cui},~\bfnm{K.}\binits{K.}} \AND
  \bauthor{\bsnm{Zhao},~\bfnm{K.}\binits{K.}}
(\byear{2008}).
\btitle{Genome-wide identification of in vivo protein-DNA binding sites from
  ChIP-seq data}.
\bjournal{Nucleic Acids Res.}
\bvolume{36}
\bpages{5221--5231}.
\end{barticle}
\endbibitem

%b21 ###
%bi21 #&#
\bibitem[\protect\citeauthoryear{Kallenberg and Ledwina}{1999}]{kallenberg1999}
\begin{barticle}[mr]
\bauthor{\bsnm{Kallenberg},~\bfnm{Wilbert C.~M.}\binits{W.~C.~M.}} \AND
  \bauthor{\bsnm{Ledwina},~\bfnm{Teresa}\binits{T.}}
(\byear{1999}).
\btitle{Data-driven rank tests for independence}.
\bjournal{J. Amer. Statist. Assoc.}
\bvolume{94}
\bpages{285--301}.
\bid{issn={0162-1459}, mr={1689233}}
\end{barticle}
\endbibitem

%b22 ###
%bi22 #&#
\bibitem[\protect\citeauthoryear{Kharchenko, Tolstorukov and Park}{2008}]{spp}
\begin{barticle}[auto:STB|2011-03-03|12:04:44]
\bauthor{\bsnm{Kharchenko},~\bfnm{P.~V.}\binits{P.~V.}},
  \bauthor{\bsnm{Tolstorukov},~\bfnm{M.~Y.}\binits{M.~Y.}} \AND
  \bauthor{\bsnm{Park},~\bfnm{P.~J.}\binits{P.~J.}}
(\byear{2008}).
\btitle{Design and analysis of ChIP-seq experiments for DNA-binding proteins}.
\bjournal{Nature Biotechnology}
\bvolume{26}
\bpages{1351--1359}.
\end{barticle}
\endbibitem

%b23 ###
%bi23 #&#
\bibitem[\protect\citeauthoryear{Kheradpour et~al.}{2007}]{kheradpour2007}
\begin{barticle}[auto:STB|2011-03-03|12:04:44]
\bauthor{\bsnm{Kheradpour},~\bfnm{P.}\binits{P.}},
  \bauthor{\bsnm{Stark},~\bfnm{A.}\binits{A.}},
  \bauthor{\bsnm{Roy},~\bfnm{S.}\binits{S.}} \AND
  \bauthor{\bsnm{Kellis},~\bfnm{M.}\binits{M.}}
(\byear{2007}).
\btitle{Reliable prediction of regulator targets using 12 drosophila genomes}.
\bjournal{Genome Res.}
\bvolume{17}
\bpages{1919--1931}.
\end{barticle}
\endbibitem

%b24 ###
%bi24 #&#
\bibitem[\protect\citeauthoryear{Kuo et~al.}{2006}]{kuo2006}
\begin{barticle}[auto:STB|2011-03-03|12:04:44]
\bauthor{\bsnm{Kuo},~\bfnm{W.}\binits{W.}},
  \bauthor{\bsnm{Liu},~\bfnm{F.}\binits{F.}},
  \bauthor{\bsnm{Trimarchi},~\bfnm{J.}\binits{J.}},
  \bauthor{\bsnm{Punzo},~\bfnm{C.}\binits{C.}},
  \bauthor{\bsnm{Lombardi},~\bfnm{M.}\binits{M.}},
  \bauthor{\bsnm{Sarang},~\bfnm{J.}\binits{J.}},
  \bauthor{\bsnm{Whipple},~\bfnm{M.~E.}\binits{M.~E.}} \betal{et~al.}
(\byear{2006}).
\btitle{A sequence-oriented comparison of gene expression measurements across
  different hybridization-based technologies}.
\bjournal{Nature Biotechnology}
\bvolume{24}
\bpages{832--840}.
\end{barticle}
\endbibitem

%b25 ###
%bi25 #&#
\bibitem[\protect\citeauthoryear{Lehmann}{2006}]{lehmann2006}
\begin{bbook}[auto:STB|2011-03-03|12:04:44]
\bauthor{\bsnm{Lehmann},~\bfnm{Erich~L.}\binits{E.~L.}}
(\byear{2006}).
\btitle{Nonparametrics: Statistical Methods Based on Ranks}, \bedition{2nd} ed.
\bpublisher{Springer}, \baddress{New York}.
\end{bbook}
\endbibitem

%b26 ###
%bi26 #&#
\bibitem[\protect\citeauthoryear{Li et~al.}{2011}]{li2011}
\begin{bmisc}[auto:STB|2011-03-03|12:04:44]
\bauthor{\bsnm{Li},~\bfnm{Q.}\binits{Q.}},
  \bauthor{\bsnm{Brown},~\bfnm{J.~B.}\binits{J.~B.}},
  \bauthor{\bsnm{Huang},~\bfnm{H.}\binits{H.}} \AND
  \bauthor{\bsnm{Bickel},~\bfnm{P.~J.}\binits{P.~J.}}
(\byear{2011}).
\bhowpublished{Supplement to ``Measuring reproducibility of high-throughput
  experiments.''
  \href{http://dx.doi.org/10.1214/11-AOAS466SUPP}{DOI:10.1214/11-AOAS466SUPP}}.
\end{bmisc}
\endbibitem

%b27 ###
%bi27 #&#
\bibitem[\protect\citeauthoryear{MAQC consortium}{2006}]{maqc2006}
\begin{bmisc}[auto:STB|2011-03-03|12:04:44]
\borganization{MAQC consortium}
(\byear{2006}).
\bhowpublished{The microarray quality control (MAQC) project shows inter- and
  intraplatform reproducibility of gene expression measurements. \textit{Nature
  Biotechnology} \textbf{24} 1151--1161}.
\end{bmisc}
\endbibitem

%b28 ###
%bi28 #&#
\bibitem[\protect\citeauthoryear{McLachlan}{1987}]{mclachlan1987}
\begin{barticle}[auto:STB|2011-03-03|12:04:44]
\bauthor{\bsnm{McLachlan},~\bfnm{G.~J.}\binits{G.~J.}}
(\byear{1987}).
\btitle{On bootstrapping the likelihood ratio test statistic for the number of
  components in a normal mixture}.
\bjournal{Applied Statistics}
\bvolume{36}
\bpages{318--324}.
\end{barticle}
\endbibitem

%b29 ###
%bi29 #&#
\bibitem[\protect\citeauthoryear{Mortazavi et~al.}{2008}]{erange}
\begin{barticle}[auto:STB|2011-03-03|12:04:44]
\bauthor{\bsnm{Mortazavi},~\bfnm{A.}\binits{A.}},
  \bauthor{\bsnm{Williams},~\bfnm{B.~A.}\binits{B.~A.}},
  \bauthor{\bsnm{McCue},~\bfnm{K.}\binits{K.}},
  \bauthor{\bsnm{Schaeffer},~\bfnm{L.}\binits{L.}} \AND
  \bauthor{\bsnm{Wold},~\bfnm{B.}\binits{B.}}
(\byear{2008}).
\btitle{Mapping and quantifying mammalian transcriptomes by RNA-seq}.
\bjournal{Nature Methods}
\bvolume{5}
\bpages{621--628}.
\end{barticle}
\endbibitem

%b33 ###
%bi30 #&#
\bibitem[\protect\citeauthoryear{Nelson}{1999}]{nelson1999}
\begin{bbook}[auto:STB|2011-03-03|12:04:44]
\bauthor{\bsnm{Nelson},~\bfnm{R.~B.}\binits{R.~B.}}
(\byear{1999}).
\btitle{An Introduction to Copulas},
\bedition{2nd} ed.
\bpublisher{Springer},
\baddress{New York}.
\end{bbook}
\endbibitem

%b30 ###
%bi31 #&#
\bibitem[\protect\citeauthoryear{Oakes}{1994}]{oakes1994}
\begin{barticle}[mr]
\bauthor{\bsnm{Oakes},~\bfnm{David}\binits{D.}}
(\byear{1994}).
\btitle{Multivariate survival distributions}.
\bjournal{J. Nonparametr. Stat.}
\bvolume{3}
\bpages{343--354}.
\bid{doi={10.1080/10485259408832593}, issn={1048-5252}, mr={1291555}}
\end{barticle}
\endbibitem

%b31 ###
%bi32 #&#
\bibitem[\protect\citeauthoryear{Park}{2009}]{park2009}
\begin{barticle}[pbm]
\bauthor{\bsnm{Park},~\bfnm{Peter~J.}\binits{P.~J.}}
(\byear{2009}).
\btitle{ChIP-seq: Advantages and challenges of a maturing technology}.
\bjournal{Nat. Rev. Genet.}
\bvolume{10}
\bpages{669--680}.
\bid{doi={10.1038/nrg2641}, issn={1471-0064}, pii={nrg2641}, pmid={19736561}}
\end{barticle}
\endbibitem

%b32 ###
%bi33 #&#
\bibitem[\protect\citeauthoryear{Rozowsky et~al.}{2009}]{peakseq}
\begin{barticle}[auto:STB|2011-03-03|12:04:44]
\bauthor{\bsnm{Rozowsky},~\bfnm{J.}\binits{J.}},
  \bauthor{\bsnm{Euskirchen},~\bfnm{G.}\binits{G.}},
  \bauthor{\bsnm{Auerbach},~\bfnm{R.~K.}\binits{R.~K.}},
  \bauthor{\bsnm{Zhang},~\bfnm{Z.~D.}\binits{Z.~D.}},
  \bauthor{\bsnm{Gibson},~\bfnm{T.}\binits{T.}},
  \bauthor{\bsnm{Bjornson},~\bfnm{R.}\binits{R.}},
  \bauthor{\bsnm{Carriero},~\bfnm{N.}\binits{N.}},
  \bauthor{\bsnm{Snyder},~\bfnm{M.}\binits{M.}} \AND
  \bauthor{\bsnm{Gerstein},~\bfnm{M.~B.}\binits{M.~B.}}
(\byear{2009}).
\btitle{PeakSeq enables systematic scoring of ChIP-seq experiments relative to
  controls}.
\bjournal{Nature Biotechnology}
\bvolume{27}
\bpages{66--75}.
\end{barticle}
\endbibitem



%b34 ###
%bi34 #&#
\bibitem[\protect\citeauthoryear{Sklar}{1959}]{sklar1959}
\begin{barticle}[mr]
\bauthor{\bsnm{Sklar},~\bfnm{M.}\binits{M.}}
(\byear{1959}).
\btitle{Fonctions de r\'epartition \`a {$n$} dimensions et leurs marges}.
\bjournal{Publ. Inst. Statist. Univ. Paris}
\bvolume{8}
\bpages{229--231}.
\bid{mr={0125600}}
\end{barticle}
\endbibitem

%b35 ###
%bi35 #&#
\bibitem[\protect\citeauthoryear{Storey}{2002}]{storey2002}
\begin{barticle}[mr]
\bauthor{\bsnm{Storey},~\bfnm{John~D.}\binits{J.~D.}}
(\byear{2002}).
\btitle{A direct approach to false discovery rates}.
\bjournal{J. R. Stat. Soc. Ser. B Stat. Methodol.}
\bvolume{64}
\bpages{479--498}.
\bid{doi={10.1111/1467-9868.00346}, issn={1369-7412}, mr={1924302}}
\end{barticle}
\endbibitem

%b36 ###
%bi36 #&#
\bibitem[\protect\citeauthoryear{Storey}{2003}]{storey2003}
\begin{barticle}[mr]
\bauthor{\bsnm{Storey},~\bfnm{John~D.}\binits{J.~D.}}
(\byear{2003}).
\btitle{The positive false discovery rate: A {B}ayesian interpretation and the
  {$q$}-value}.
\bjournal{Ann. Statist.}
\bvolume{31}
\bpages{2013--2035}.
\bid{doi={10.1214/aos/1074290335}, issn={0090-5364}, mr={2036398}}
\end{barticle}
\endbibitem

%b37 ###
%bi37 #&#
\bibitem[\protect\citeauthoryear{Stouffer et~al.}{1949}]{stouffer1949}
\begin{bmisc}[auto:STB|2011-03-03|12:04:44]
\bauthor{\bsnm{Stouffer},~\bfnm{S.~A.}\binits{S.~A.}},
  \bauthor{\bsnm{Suchman},~\bfnm{E.~A.}\binits{E.~A.}},
  \bauthor{\bsnm{DeVinney},~\bfnm{L.~C.}\binits{L.~C.}},
  \bauthor{\bsnm{Star},~\bfnm{S.~A.}\binits{S.~A.}} \AND
  \bauthor{\bsnm{Williams},~\bfnm{J.}\binits{J.}}
(\byear{1949}).
\bhowpublished{\textit{The American Soldier: Vol. 1. Adjustment During Army
  Life.} Princeton Univ. Press,
  Princeton, NJ}.
\end{bmisc}
\endbibitem

%b38 ###
%bi38 #&#
\bibitem[\protect\citeauthoryear{Sun and Cai}{2007}]{sun2007}
\begin{barticle}[mr]
\bauthor{\bsnm{Sun},~\bfnm{Wenguang}\binits{W.}} \AND
  \bauthor{\bsnm{Cai},~\bfnm{T.~Tony}\binits{T.~T.}}
(\byear{2007}).
\btitle{Oracle and adaptive compound decision rules for false discovery rate
  control}.
\bjournal{J. Amer. Statist. Assoc.}
\bvolume{102}
\bpages{901--912}.
\bid{doi={10.1198/016214507000000545}, issn={0162-1459}, mr={2411657}}
\end{barticle}
\endbibitem

%b39 ###
%bi39 #&#
\bibitem[\protect\citeauthoryear{Thurman et~al.}{2011}]{hotspot}
\begin{bmisc}[auto:STB|2011-03-03|12:04:44]
\bauthor{\bsnm{Thurman},~\bfnm{R.}\binits{R.}},
  \bauthor{\bsnm{Hawrylycz},~\bfnm{M.}\binits{M.}},
  \bauthor{\bsnm{Kuehn},~\bfnm{S.}\binits{S.}},
  \bauthor{\bsnm{Haugen},~\bfnm{E.}\binits{E.}} \AND
  \bauthor{\bsnm{Stamatoyannopoulos},~\bfnm{S.}\binits{S.}}
(\byear{2011}).
\bhowpublished{Hotspot: A scan statistic for identifying enriched regions of
  short-read sequence tags. Unpublished manuscript,
  Univ. Washington}.
\end{bmisc}
\endbibitem

%b40 ###
%bi40 #&#
\bibitem[\protect\citeauthoryear{Valouev et~al.}{2008}]{quest}
\begin{barticle}[auto:STB|2011-03-03|12:04:44]
\bauthor{\bsnm{Valouev},~\bfnm{A.}\binits{A.}},
  \bauthor{\bsnm{Johnson},~\bfnm{D.~S.}\binits{D.~S.}},
  \bauthor{\bsnm{Sundquist},~\bfnm{A.}\binits{A.}},
  \bauthor{\bsnm{Medina},~\bfnm{C.}\binits{C.}},
  \bauthor{\bsnm{Anton},~\bfnm{E.}\binits{E.}},
  \bauthor{\bsnm{Batzoglou},~\bfnm{S.}\binits{S.}},
  \bauthor{\bsnm{Myers},~\bfnm{R.~M.}\binits{R.~M.}} \AND
  \bauthor{\bsnm{Sidow},~\bfnm{A.}\binits{A.}}
(\byear{2008}).
\btitle{Genome-wide analysis of transcription factor binding sites based on
  ChIP-seq data}.
\bjournal{Nature Methods}
\bvolume{5}
\bpages{829--834}.
\end{barticle}
\endbibitem

%b41 ###
%bi41 #&#
\bibitem[\protect\citeauthoryear{Zhang et~al.}{2008}]{macs}
\begin{barticle}[auto:STB|2011-03-03|12:04:44]
\bauthor{\bsnm{Zhang},~\bfnm{Y.}\binits{Y.}},
  \bauthor{\bsnm{Liu},~\bfnm{T.}\binits{T.}},
  \bauthor{\bsnm{Meyer},~\bfnm{C.~A.}\binits{C.~A.}},
  \bauthor{\bsnm{Eeckhoute},~\bfnm{J.}\binits{J.}},
  \bauthor{\bsnm{Johnson},~\bfnm{D.~S.}\binits{D.~S.}},
  \bauthor{\bsnm{Bernstein},~\bfnm{B.~E.}\binits{B.~E.}},
  \bauthor{\bsnm{Nussbaum},~\bfnm{C.}\binits{C.}},
  \bauthor{\bsnm{Myers},~\bfnm{R.~M.}\binits{R.~M.}},
  \bauthor{\bsnm{Brown},~\bfnm{M.}\binits{M.}},
  \bauthor{\bsnm{Li},~\bfnm{W.}\binits{W.}} \AND
  \bauthor{\bsnm{Liu},~\bfnm{X.~S.}\binits{X.~S.}}
(\byear{2008}).
\btitle{Model-based analysis of ChIP-seq (MACS)}.
\bjournal{Genome Biology}
\bvolume{9}
\bpages{R137}.
\end{barticle}
\endbibitem

\end{thebibliography}
\end{document}